\documentclass[twoside]{article}
\usepackage{amsfonts,amssymb,amsbsy,textcomp,marvosym,picins,amsmath,bm,caption,threeparttable,amsthm,subfigure}
\usepackage{eurosym,mathrsfs,fancyhdr,CJK,multicol,multirow,graphics,indentfirst,color,bm,upgreek,booktabs,graphicx}
\usepackage{epstopdf}
\usepackage[ruled,linesnumbered,vlined]{algorithm2e}
\usepackage{algpseudocode}
\usepackage{comment}
\usepackage{bbm}
\usepackage{textcomp}
\usepackage{xcolor}
\usepackage{array}
\usepackage{hyperref}
\usepackage{capt-of}
\usepackage{authblk}
\theoremstyle{definition}
\newtheorem{definition}{Definition}
\def\footnoterule{\kern 1mm \hrule width 10cm \kern 2mm}

\allowdisplaybreaks
\catcode`@=11
\def\title#1{\vspace{3mm}\begin{flushleft}\vglue-.1cm\Large\bf\boldmath\protect\baselineskip=18pt plus.2pt minus.1pt #1
\end{flushleft}\vspace{1mm} }
\def\author#1{\begin{flushleft}\normalsize #1\end{flushleft}\vspace*{-4pt} \vspace{3mm}}
\def\address#1#2{\begin{flushleft}\vglue-.35cm${}^{#1}$\small\it #2\vglue-.35cm\end{flushleft}\vspace{-2mm}\par}

\catcode`@=11
\def\section{\@startsection{section}{1}{\z@}%
 {-3ex \@plus -.3ex \@minus -.2ex}%
 {2.2ex \@plus.2ex}%
{\normalfont\normalsize\protect\baselineskip=14.5pt plus.2pt minus.2pt\bfseries}}
\def\subsection{\@startsection{subsection}{2}{\z@}%
 {-3ex\@plus -.2ex \@minus -.2ex}%
 {2ex \@plus.2ex}%
{\normalfont\normalsize\protect\baselineskip=12.5pt plus.2pt minus.2pt\bfseries}}
\def\subsubsection{\@startsection{subsubsection}{3}{\z@}%
 {-2.2ex\@plus -.21ex \@minus -.2ex}%
 {1.4ex \@plus.2ex}
{\normalfont\normalsize\protect\baselineskip=12pt plus.2pt minus.2pt\sl}}

\begin{document}
\begin{CJK*}{GBK}{song}
\thispagestyle{empty}
\vspace*{-13mm}
\noindent {\small First Author, Second Author, Third
Author {\it et al.} Journal of computer science and technology: Instruction for authors.
JOURNAL OF COMPUTER SCIENCE AND TECHNOLOGY \ 33(1): \thepage--\pageref{last-page}
\ January 2018. DOI 10.1007/s11390-015-0000-0}
\vspace*{2mm}

\title{Extracting Variable-Depth Logical Document Hierarchy from Long Documents: Method, Evaluation, and Application}

\author{Rong-Yu Cao$^{1,2}$, Yi-Xuan Cao$^{1,2}$, Gan-Bin Zhou$^{3}$, Ping Luo$^{1,2,4}$\\}

\address{1}{Key Laboratory of Intelligent Information Processing of Chinese Academy of Sciences, Institute of Computing Technology, Chinese Academy of Sciences, Beijing 100190, China}
\address{2}{University of Chinese Academy of Sciences, Beijing 100049, China}
\address{3}{WeChat Search Application Department, Tencent Holdings Ltd., Beijing 100080, China}
\address{4}{Peng Cheng Laboratory, Shenzhen 518066, China}

\vspace{2mm}

\noindent E-mail: caorongyu19b@ict.ac.cn; caoyixuan@ict.ac.cn; ganbinzhou@tencent.com; luop@ict.ac.cn\\[-1mm]

\noindent Received February 25, 2021; accepted March 03, 2021.\\[1mm]

\let\thefootnote\relax\footnotetext{{}\\[-4mm]\indent\ Regular Paper\\[.5mm]
\indent\ This work was supported by the National Key Research and Development Program of China under Grant No. 2017YFB1002104, the National Natural Science Foundation of China under Grant Nos. 62076231 and U1811461. \\[.5mm]
\\[.5mm]\indent\ \copyright Institute of Computing Technology, Chinese Academy of Sciences 2021}

\noindent {\small\bf Abstract} \quad  {\small
In this paper, we study the problem of extracting variable-depth ``logical document hierarchy'' from long documents, namely organizing the recognized ``physical document objects'' into hierarchical structures.
The discovery of logical document hierarchy is the vital step to support many downstream applications (e.g. passage-based retrieval and high-quality information extraction).
However, long documents, containing hundreds or even thousands of pages and variable-depth hierarchy, challenge the existing methods.
To address these challenges, we develop a framework, namely Hierarchy Extraction from Long Document (HELD), where we ``sequentially'' insert each physical object at the proper on of the current tree.
Determining whether each possible position is proper or not can be formulated as a binary classification problem.
To further improve its effectiveness and efficiency, we study the design variants in HELD, including traversal orders of the insertion positions, heading extraction explicitly or implicitly, tolerance to insertion errors in predecessor steps, and so on.
As for evaluations, we find that previous studies ignore the error that the depth of a node is correct while its path to the root is wrong.
Since such mistakes may worsen the downstream applications seriously, a new measure is developed for more careful evaluation.
The empirical experiments based on thousands of long documents from Chinese, English financial market and English scientific publication show that the HELD model with the ``root-to-leaf'' traversal order and explicit heading extraction is the best choice to achieve the tradeoff between effectiveness and efficiency with the accuracy of 0.9726, 0.7291 and 0.9578 in Chinese financial, English financial and arXiv datasets, respectively.
Finally, we show that logical document hierarchy can be employed to significantly improve the performance of the downstream passage retrieval task.
In summary, we conduct a systematic study on this task in terms of methods, evaluations, and applications.}

\vspace*{2mm}


\vspace*{2mm}

\end{CJK*}
\baselineskip=18pt plus.2pt minus.2pt
\parskip=0pt plus.2pt minus0.2pt
\begin{multicols}{2}

\section{Introduction}
\label{sec:intro}

\begin{figure*}[!htb]
    \centering
    \subfigure[]{\label{fig:intro_diff_a_b}
    \includegraphics[width=2.5in]{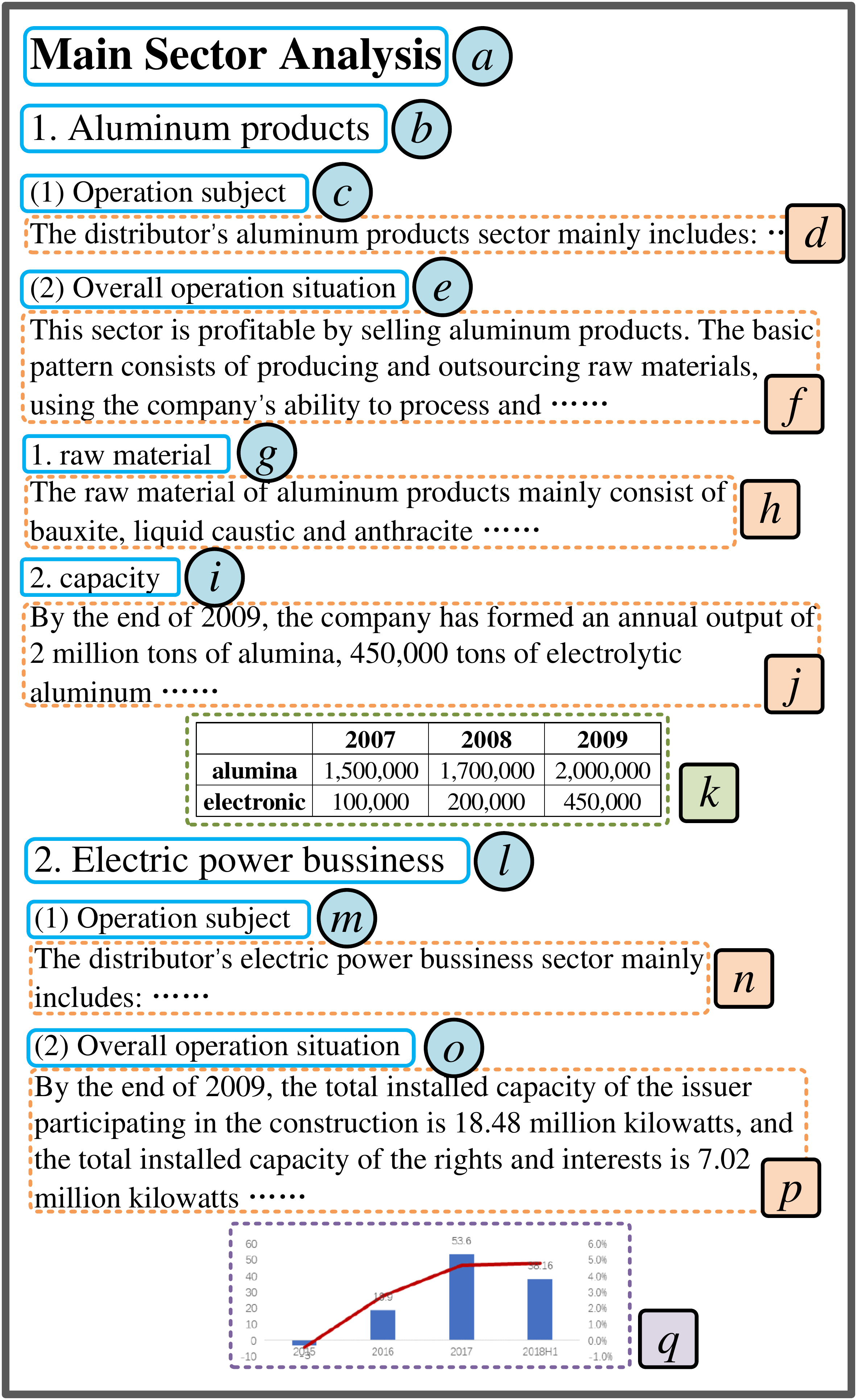}}
    \hspace{10mm}
    \subfigure[]{\label{fig:intro_diff_a_a}
    \includegraphics[width=2.15in]{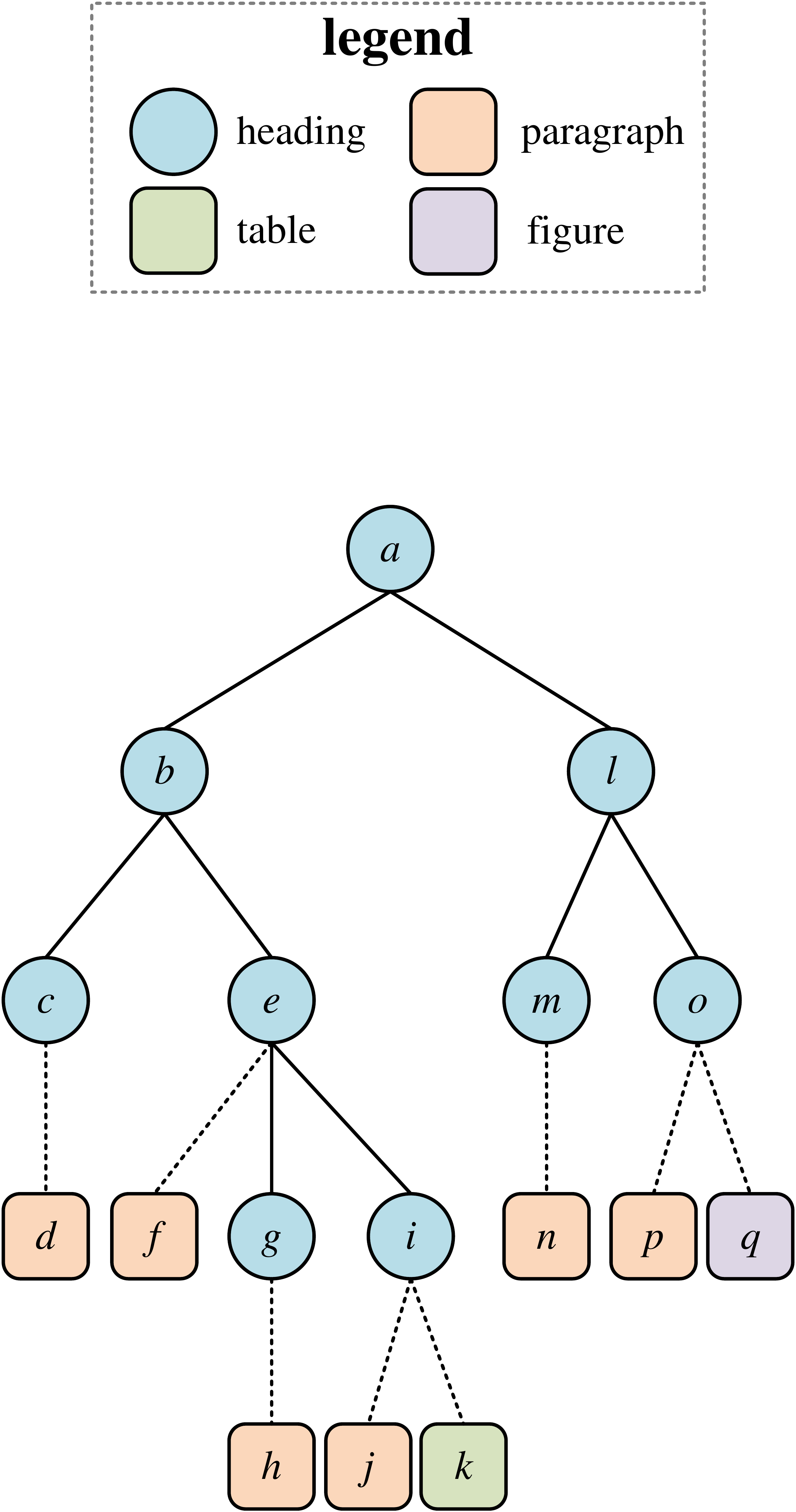}}
    \captionof{figure}{An example about logical document hierarchy discovery. (a) The example document page and the physical objects on this page. (b) The logical document hierarchy of this document page.}\label{fig:intro_diff_a}
\end{figure*}

Recently, the amount of electronic documents have increased rapidly along with the IT penetration into various vertical domains, such as financial, legal, government and education fields. To gain valuable insights from these unstructured documents, it is of the highest importance to obtain their underlying document structures so that these documents can be reedited, restyled, or reflowed to support many downstream natural language processing (NLP) and text mining applications. However, the transformation from the editing formats (e.g. WORD and LaTeX) of these documents to their display formats (e.g. PDF and JPG) only guarantees the appropriateness of document layout, while their underlying physical and logical structures are either partially or even completely lost~\cite{bloechle2010physical}. Hence, it is still an open issue to make this transformation reversible generally.

\begin{figure*}[!htb]
    \centering
    \subfigure[]{\label{fig:intro_page_dist}
    \includegraphics[width=3.2in]{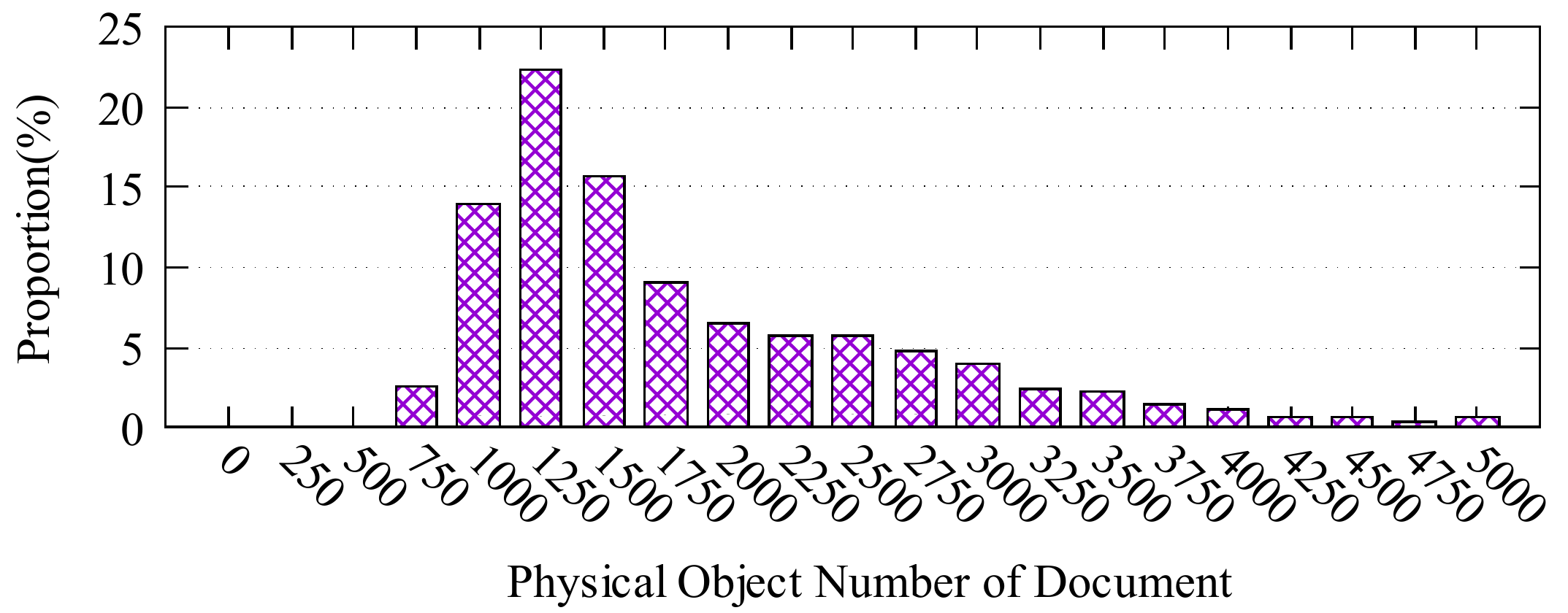}}
    \subfigure[]{\label{fig:intro_level_dist}
    \includegraphics[width=3.2in]{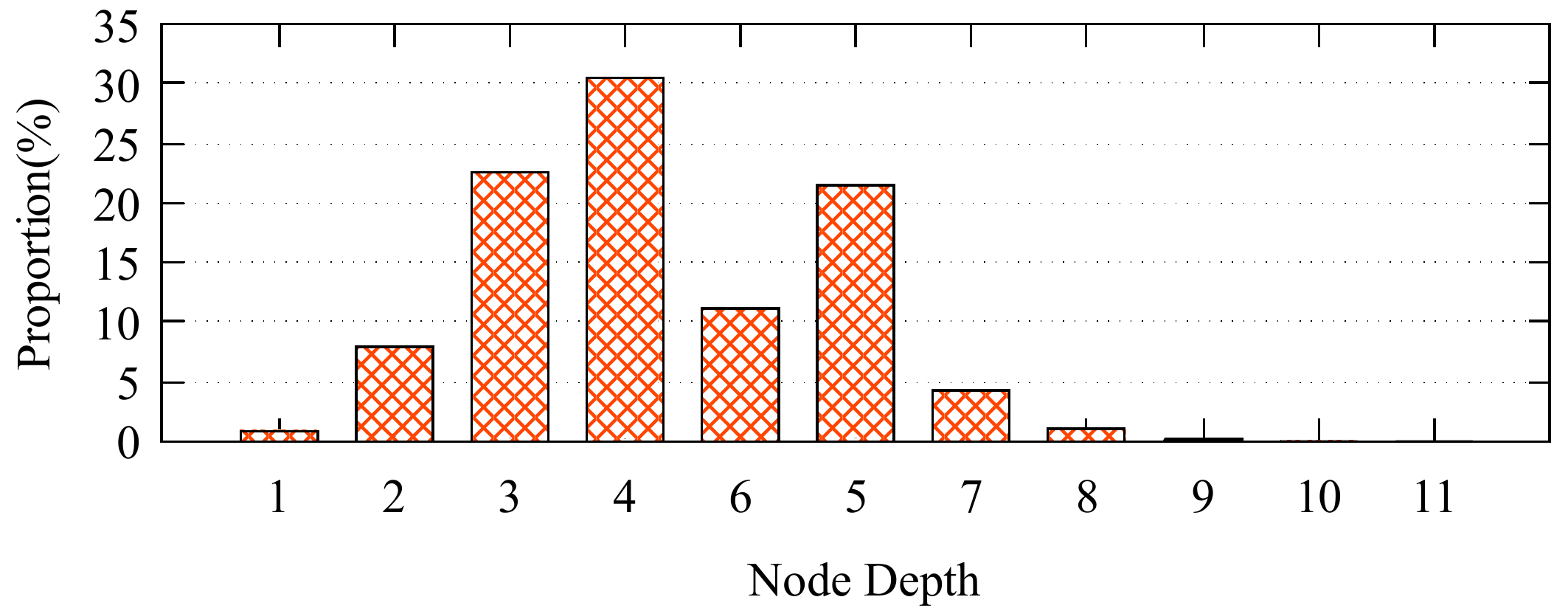}}
    \caption{The distribution on the benchmark documents. (a) Distribution of physical object number. (b) Distribution of headings on each depth}
    \label{fig:intro_distribution}
\end{figure*}

To this end, in this paper, we study the problem of extracting variable-depth logical document hierarchy from long documents, which aims to organize the recognized ``physical document objects'' into hierarchical structures. A typical example of this task with a one-page document and its logical document hierarchy is shown in Fig.~\ref{fig:intro_diff_a}. Here, physical objects refer to paragraphs, tables, charts and figures in a document ~\cite{Mao2003Document}. We assume that a predecessor step detects these objects and ranks them by the reading order already. The goal of this study is to transform the flat structure of these physical objects into a hierarchical structure, which reflects the parallel and containment relationship between these physical objects. The discovery of logical document hierarchy helps to support many downstream applications such as hierarchical browsing, passage-based retrieval, high-quality information extraction and reading comprehension~\cite{pembe2007heading,geva2018learning,turtle1991inference,summers1998automatic}.

Although the recovery of logical document hierarchy attracts extensive researches~\cite{Luong2010Logical,Pembe2010,Ramakrishnan2012Layout,Manabe2015,Rahman2017Understanding,Bentabet2019}, most studies focus on scientific papers or web pages, where only tens of pages are contained and the logical hierarchy is often fixed and shallow (4 levels at the most). Recently, millions of disclosure documents in the financial area from different countries are published every year. However, these documents, such as annual reports, prospectuses, etc., usually have hundreds of pages, and their hierarchies are much deeper with variable depth. Based on the thousands of benchmark documents with their annotated hierarchies, Fig.~\ref{fig:intro_distribution} shows the distribution of physical object numbers and the distribution of headings on each depth. We observe that all the documents have at least 500 physical objects, 90\% of headings locate on the 3rd to 7th level of the trees, and the maximal depth is 11.

Such variable-depth logical hierarchy from long document challenges the existing methods on logical hierarchy recovery~\cite{Pembe2010,Manabe2015,Rahman2017Understanding,Bentabet2019}. Previous solutions can be grouped into three types. The first type~\cite{Rahman2017Understanding,Bentabet2019} formulates this task as a sequence labeling task, which employs Long Short Term Memory (LSTM) or Conditional Random Field (CRF) to extract contextual features of surrounding physical objects and classifies each heading into the absolute hierarchical depth. However, this type of method fixes the space of depth labels and assigns an ``absolute'' depth to each physical object. In this study, we argue that since the hierarchical depth of physical objects depends on the containment and parallel relationship between contextual physical objects, the hierarchical depth should be considered as a ``relative'' concept than an ``absolute'' one. Additionally, due to extremely long distances among physical objects in the documents with hundreds of pages, sequence labeling based methods might not work well in capturing such long-distance context. The experimental results in this study also show that these methods obtain lower accuracy on our benchmark documents. The second type of method in~\cite{Manabe2015,Conway1993Parser} is the rule-based method. They mostly propose some assumptions on logical hierarchy. For example, the study in~\cite{Manabe2015} assumes that the headings with the same visual and textual style always locate at the same hierarchical depth. However, as illustrated by our benchmark documents, these assumptions are not always true. The third type is the hierarchy generation based method in~\cite{Pembe2010}. It dynamically generates the logical hierarchy by considering the containment and parallel relationship between headings. However, this work lacks systematic studies on the possible variants of the generation process.

Inspired by how humans construct hierarchical trees in reading, we propose a novel model, namely Hierarchy Extraction from Long Document (HELD). Specifically, we sequentially insert each physical object at the proper position of the tree. By a certain traversal order, we inquire about all the possible insertion positions in the current tree until we find the proper one. Determining whether each possible position is proper or not can be formulated as a binary classification problem, namely the ``put-or-skip'' module. The hierarchical tree is generated until all the physical objects have been inserted. In this framework, the put-or-skip module is the key step. We propose an LSTM based sub-model to detect the relative containment and parallel relationships between physical objects. The combination of both visual and textual features is adopted to capture the relationship between the local context of each insertion position and the physical object to be inserted. Furthermore, we study the design variants in HELD, including traversal orders of the insertion positions, heading extraction explicitly or implicitly, tolerance to insertion errors in predecessor steps, and so on.

As for evaluations, we find that previous studies ignore the error that the depth of a node is correct while its path to the root is wrong. Since such mistakes might seriously worsen the downstream applications, we propose a new measure, where an inserted node is correct if and only the path from the root to itself is completely the same as the ground-truth path. We argue that this measure should be adopted in future studies of logical document hierarchy discovery.

Based on 1030 Chinese documents from the financial domain (namely the Chinese dataset), 1203 English documents from the financial domain (namely the English dataset) and 1732 arXiv documents from the scientific domain (namely the arXiv dataset), we compare the proposed HELD model with the rule-based, sequence-tagging and existing generation-based methods. In the Chinese dataset, the HELD model achieves the best accuracy of 0.9731, while the rule-based, sequence-tagging and existing generation-based methods obtain the accuracy of 0.3764, 0.9403 and 0.9339 respectively. In the English dataset, the HELD model achieves the best accuracy of 0.7301, while the three baseline methods obtain the accuracy of 0.4779, 0.6436 and 0.6563, respectively. In the arXiv dataset, the HELD model achieves the best accuracy of 0.9578, while the three baseline methods obtain the accuracy of 0.8375, 0.8908 and 0.9034, respectively. To achieve the tradeoff between effectiveness and efficiency, the HELD model with the ``root-to-leaf'' traversal order and explicit heading extraction is the best choice with 0.9726, 0.7291 and 0.9567 accuracy on Chinese, English and arXiv datasets, respectively, and 8.3x speedup in inference efficiency.

Finally, we demonstrate that logical document hierarchy can be employed to significantly improve the performance of a downstream application, namely passage retrieval in a long document.

In summary, we conduct a systematic study on extracting variable-depth logical document hierarchy from long documents in terms of methods, evaluations, and applications.

This paper has the following organization. In Section 2, we review some related work.
Details of the proposed HELD model are described in Section 3. Section 4 and Section 5 present the configurations and results of experiments. Section 6 introduces a downstream application - passage-based retrieval. This paper ends with a summary and a brief discussion of future work.

\section{Related Work}

\subsection{Logical Document Hierarchy Extraction}
\label{sec:relatedwork}

The discovery of logical document hierarchy is a conventional task, Summers {\it et al.}~\cite{summers1998automatic} gave a proper definition: logical structure consists of a hierarchy of segments of the document, each of which corresponds to a visually distinguished semantic component of the document. Generally, previous studies can be grouped into the rule-based method and learning-based method.

For rule-based methods, Tsujimoto {\it et al.}~\cite{tsujimoto1990understanding} aimed to discover logical structure in multi-article newspapers, by using some generic transformation rules and a virtual field separator technique. Conway {\it et al.}~\cite{Conway1993Parser} used a set of grammar rules, which is a string of components specified by neighbor relation, and page parsing techniques to recognize document logical structure. Manabe {\it et al.}~\cite{Manabe2015} proposed some assumptions, like two headings with the same visual style should locate at the same significant level. Then, they sort these headings by some visual styles (e.g. font size, bold or italic) and then generate a heading hierarchy.

Learning-based methods can be further separated into two classes, sequence labeling-based and tree generation-based methods. Some of sequence labeling-based methods first recognize physical objects and determine the reading order of them, then use different models to classify the absolute hierarchical depth of each physical object. For example, Luong {\it et al.}~\cite{Luong2010Logical} used conditional random field (CRF), Rahman {\it et al.}~\cite{Rahman2017Understanding} used RNNs and Bentabet {\it et al.}~\cite{Bentabet2019} used LSTMs to classify physical into four categories: main-text, section-header, subsection-header and subsubsection-header. Other work~\cite{Constantin2013PDFX,Tkaczyk2015CERMINE} combine rule-based and model-based method to extract logical structure. For tree generation-based method, Pembe {\it et al.}~\cite{Pembe2010} proposed a tree-based learning approach to generate logical hierarchy node by node, by considering the containment and parallel relationship between nodes.

\subsection{Physical Structure Recognition}

Physical structure recognition is a basic step for extracting logical document hierarchy. It focuses on dividing the document into flat segmentations, rather than a hierarchy~\cite{summers1995toward}. Here, flat segmentation represents an ordered list of physical objects (e.g. tables, paragraphs, figures, etc.~\cite{Mao2003Document}). Physical structure recognition can be categorized into top-down and bottom-up approaches.

The top-down approach~\cite{tsujimoto1990understanding,baird1990image,Nagy1992A} starts from the whole document and split it into smaller components iteratively. Tsujimoto {\it et al.}~\cite{tsujimoto1990understanding} divided the document page into some rectangle blocks and used simple rules to classify each block. Nagy {\it et al.}~\cite{Nagy1992A} proposed a vertical and horizontal cut-off-lines-based and Baird {\it et al.}~\cite{baird1990image} proposed a shape-directed-covers-based method to recursively split the document page into smaller regions.

The Bottom-up approach gathers pixels or characters into text lines then combines them into physical objects. Early work~\cite{Conway1993Parser,kopec1994document} are grammar-based methods, which design different layout grammars to analyze the physical structure. Later work consider the task as semantic segmentation or sequence tagging problem. By regarding as the semantic segmentation problem, some work~\cite{Xiao2017Learning,augusto2017fast} detect the contour of each physical object (by length algorithm~\cite{wong1982document}) and classify pixels in it (by FCN, VGG~\cite{simonyan2014very,noh2015learning}) to determine the type of physical objects. By regarding as the sequence tagging problem, some work~\cite{Luong2010Logical,he2017multi} split the document into an ordered list of text lines and determine the reading order of these text lines. Then, different methods (e.g. CRF, RNNs) are used to classify the type of each text line. Neighboring text lines with the same type will be grouped into a physical object.

\section{Hierarchy Extraction from Long Document Model}

\begin{figure*}
    \centering
    \subfigure[]{
    \includegraphics[width=5.0in]{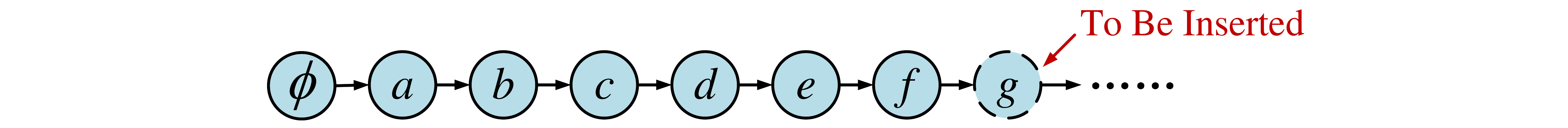}}
    \subfigure[]{
    \includegraphics[width=2.2in]{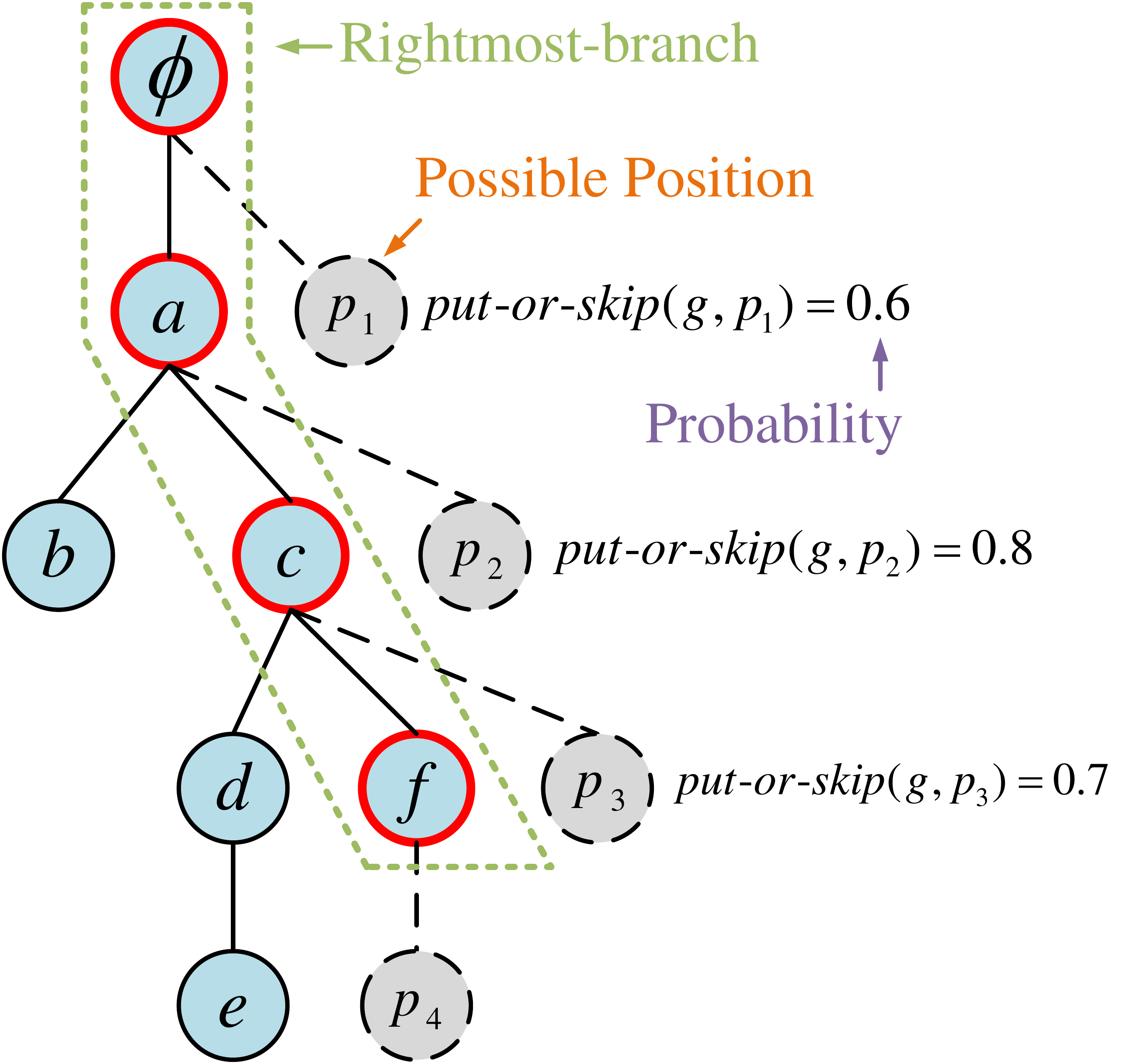}}
    \subfigure[]{
    \includegraphics[width=3.0in]{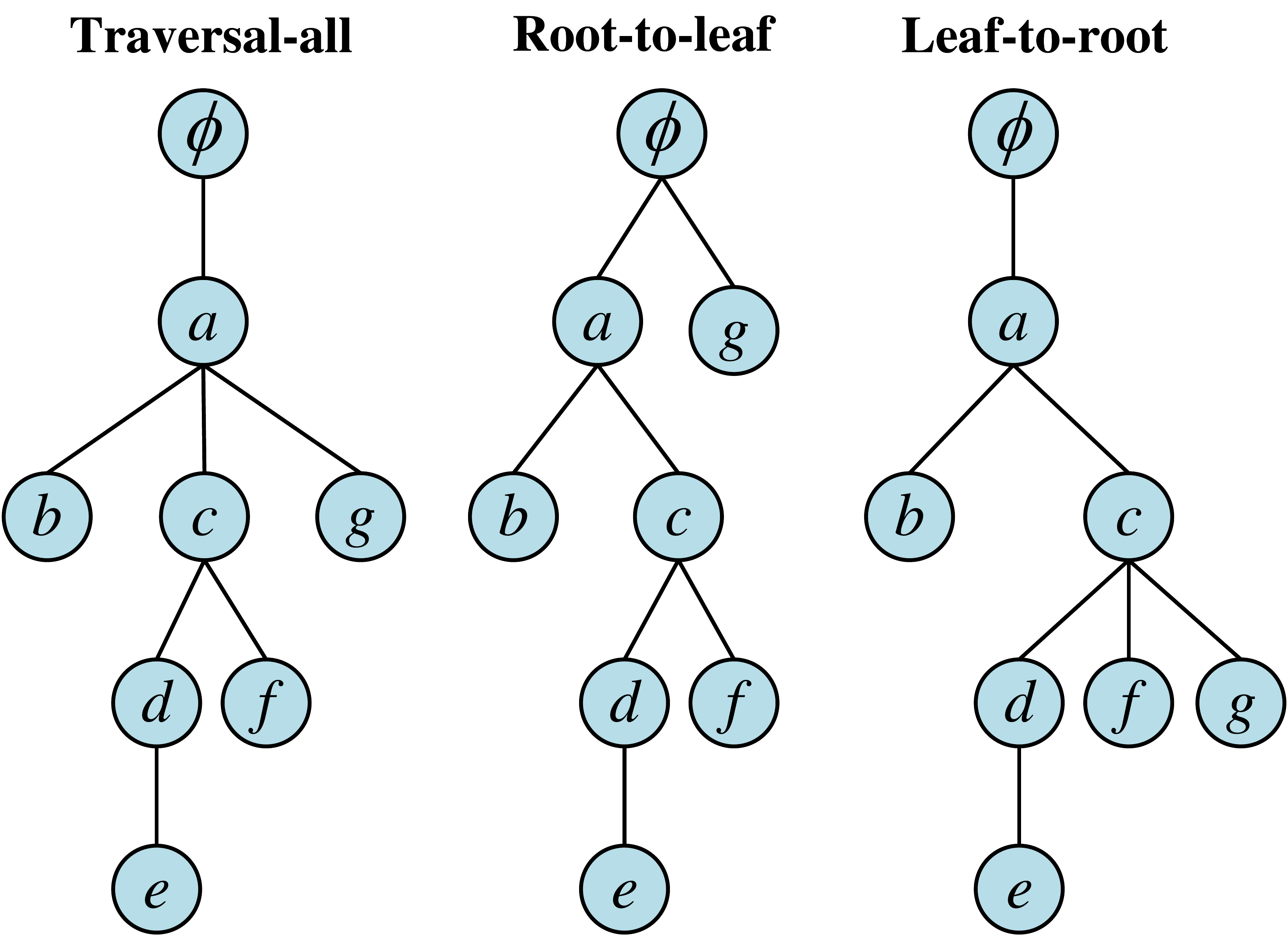}}
    \caption{Example about tree generation. (a) Inserting physical object $g$, the rightmost-branch, is $\phi$, $a$, $c$ and $f$. (b) Hierarchical tree when inserting $g$. $p_1$, $p_2$, $p_3$ and $p_4$ are possible positions for $g$ to be inserted and each has a probability. (c) The results after inserting $g$ under three traversal methods.}\label{fig:generation_process}
\end{figure*}

Here, we introduce Hierarchy Extraction from Long Document (HELD) model to convert an ordered list of physical objects (e.g. paragraphs, tables, charts and figures), namely $C=\{c_i\}_{i=1}^N$ where $N$ is the total number of physical objects, into a hierarchical tree $T$.
We assume that the list $C$ is obtained in a predecessor step - physical structure recognition.
Hence, after a careful evaluation, we adopt a commercial product, PDFLux\footnote{\url{https://pdflux.com/}. Last visited in 2021/04/28.} for this step.
It can obtain physical structures and determine a reading order on various financial documents with high accuracy, especially for disclosed financial documents.

\subsection{Framework of Hierarchical Tree Generation Sub-model}\label{sec:3.1}

When a human reads a document, in her/his mind she/he actually constructs the logical document hierarchy gradually along the sequential reading process. When encountering a physical object in the document, she makes the decision on inserting this node into one of the possible positions in the current tree. Inspired by this human process, we propose the framework of Hierarchy Extraction from Long Document (HELD). Specifically, HELD sequentially checks each physical object in $C$ one by one and inserts it into a proper position of the current tree. The key to this process is to clearly define all the possible insertion positions of the current tree. Before that, we first define the rightmost-branch of the current tree as follow.

\begin{definition}
{\it The rightmost-branch of the tree is an ordered list of nodes, where the first node is root $\phi$, and each next node is the rightmost child of the previous node.}
\end{definition}

For example, as shown in Fig.~\ref{fig:generation_process}, for the current tree its rightmost-branch is ``$\phi$, $a$, $c$, $f$'', where each node is highlighted with red circle.

With this rightmost-branch of the tree, we further define the possible insertion positions of the current tree as follow.

\begin{definition}
{\it For a new node to insert, its possible insertion positions are all the last children of the nodes in the rightmost-branch of the current tree.}
\end{definition}

As shown in Fig.~\ref{fig:generation_process}, there are four nodes in the rightmost-branch, thus there are four possible insertion positions: $p_1$, $p_2$, $p_3$, $p_4$ are the last children of $\phi$, $a$, $c$, $f$, respectively.

The correctness of this definition of possible insertion positions is guaranteed by the following theoretical analysis. It is clear that the pre-order traverse of the document hierarchical tree generates the list of physical objects in the reading order of the document. This definition can theoretically guarantee that the node to insert is always ranked at the last position of the pre-order traverse of the tree after insertion only if it is inserted into any one of these possible positions.

With all these possible positions, we need a module to decide which position is proper for the current node. Next, Subsection~\ref{sec:3.2} gives the training objective function of the whole framework, and Subsection~\ref{sec:3.3} details the module of selecting the right insertion position.

\subsection{Objective Function} \label{sec:3.2}

For each document with the physical nodes of $\{c_i\}_{i=1}^N$, the joint probability of the tree can be decomposed into the probability of each physical object $c_i$, in the condition that its previous physical objects have been already inserted. Specifically, it can be represented as follows,

\begin{equation*}
    \log P(\mathcal{T})=\sum_{i=1}^{N}\log P(c_i|\mathcal{T}_{c_{1}, c_{2}, \cdots, c_{i-1}}),
\end{equation*}
where $\mathcal{T}_{c_{1}, c_{2}, \cdots, c_{i-1}}$ is the sub-tree constructed by the nodes of $c_{1}, c_{2}, \cdots, c_{i-1}$.

In the sub-tree $\mathcal{T}_{c_{1}, c_{2}, \cdots, c_{i-1}}$, we only consider all the possible insertion positions, denoted as $S_i = \{s_i^j\}_{j=1}^{M_i}$ where $s_i^j$ is the $j$-th possible insertion position, and $M_i$ represents the number of possible insertion positions. Among these positions, we denote $s_i^*$ as the correct insertion position.

Then,  $\log P(c_i|\mathcal{T}_{c_{1}, c_{2}, \cdots, c_{i-1}})$ can be further expanded as
\begin{equation*}
    \begin{aligned}
        \sum_{j=1}^{M_i} & \left( \mathbbm{1}\left( s_i^j = s_i^* \right) \times \log P \left( c_i \ | ctx(s_i^j) \right) \right. \\
        & \left. - \mathbbm{1}\left( s_i^j \neq s_i^* \right) \times \log P \left( c_i \ | ctx(s_i^j) \right) \right),
    \end{aligned}
\end{equation*}
where $P \left( c_i \ | ctx(s_i^j) \right) $ stands for the probability of inserting node $c_i$ into $s_i^j$ and $ctx(s_i^j)$ represents the contextual information of position $s_i^j$. $\mathbbm{1}(\cdot)$ is the indicator function, which equals to 1 if the condition holds otherwise equals to 0.

Then, for a corpus of $L$ trees $\{\mathcal{T}_1,\cdots,\mathcal{T}_L\}$ we aim to maximize the following objective function

\begin{equation*}
    \begin{aligned}
        \sum_{k=1}^{L} \sum_{i=1}^{N_k} \sum_{j=1}^{M_i} & \left( \mathbbm{1}\left( s_i^j = s_i^* \right) \times \log P \left( c_i \ | ctx(s_i^j) \right) \right. \\
        & \left. - \mathbbm{1}\left( s_i^j \neq s_i^* \right) \times \log P \left( c_i \ | ctx(s_i^j) \right) \right),
    \end{aligned}
\end{equation*}
where $N_k$ is the number of physical nodes in the $k$-th tree $\mathcal{T}_k$. In this objective function, we aim to maximize $P \left( c_i \ | ctx(s_i^j) \right) $ when $s_i^j$ is the true position of $c_i$. Otherwise, we aim to minimize it.

With any annotated tree $\mathcal{T}$, it is easy to transform it into the labeled data for training $P \left( c_i \ | ctx(s_i^j) \right) $. Specifically, for each physical object $c_i$, we find out each possible insertion position $s_i^j$, and get all the corresponding tuples $<c_i, ctx(s_i^j), l_i^j>$, where $l_i^j$ equals to 1 if position $s_i^j$ is the correct position of $c_i$; otherwise it equals to 0. In this way, we can build a huge set of such tuples from all the annotated trees to train the parameters in $P \left( c_i \ | ctx(s_i^j) \right) $.

Note that all these training data are generated with the assumption that when inserting $c_i$, all the nodes ${c_{1}, c_{2}, \cdots, c_{i-1}}$ before $c_i$ are all correctly inserted. However, this is not always true in the inference process of a new document. In Subsection~\ref{sec:tolerance}, we will show how the training data can be enriched to be tolerant to some insertion errors in the predecessor steps.

\subsection{Put-Or-Skip Module}\label{sec:3.3}

Next, we will present how the put-or-skip module is built. It aims to estimate the probability of inserting a physical object $c$ into a possible insertion position $s$, denoted as $P(c|ctx(s))$. It can be regarded as a binary classification problem. Here, $ctx(s)$ refers to the contextual information of the position $s$. As shown in Fig.~\ref{fig:put_or_skip}, this context may include the siblings of $s$, and its immediate parent. We observe that this local context provides vital cues for this classification. Specifically, if $s$ is the right position for $c$, the siblings of $s$, namely $ g_1, g_2, \cdots, g_K$, might have the same format features and consecutive item number with $c$, and its immediate parent $z$ might have more prominent format style than $c$. Thus, the module is required to consider all the textual and visual features inside the local context.

\vspace{4mm}
\begin{center}
    \includegraphics[width=3.2in]{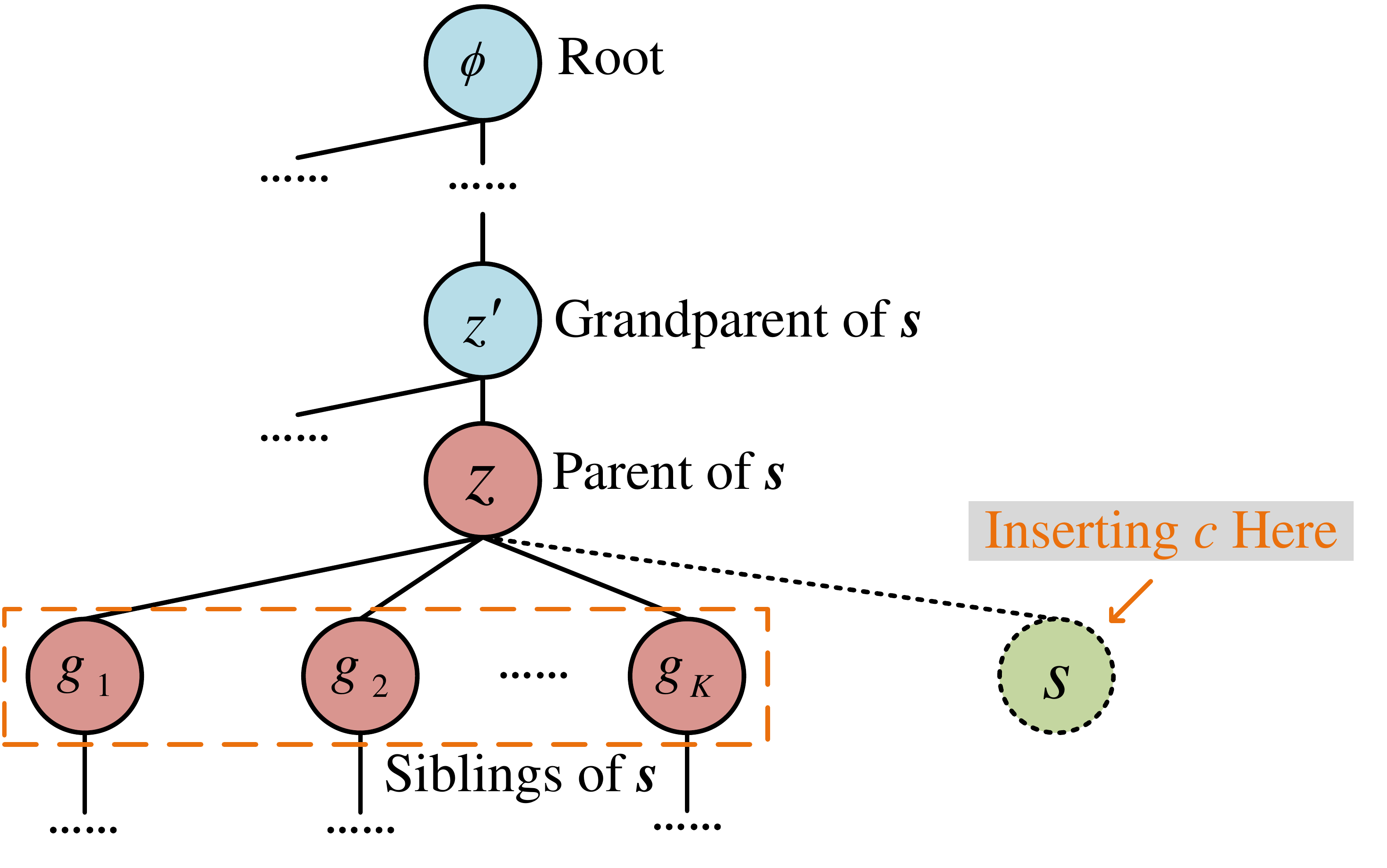}
    \vspace{1mm}
    \captionof{figure}{The context of determining whether insert physical object $c$ into position $s$.}\label{fig:put_or_skip}
\end{center}
\vspace{2mm}

To capture the textual features, we use a Bi-LSTM~\cite{schuster1997bidirectional} to model the text inside each physical object in the context. Specifically, it successively receives every word in the text of each physical object $x$, and outputs a fixed-length vector, namely $ \bm{v}_x$. Thus, we can obtain text representation of $c, g_1, \cdots, g_K$ and $z$, namely $\bm{v}_c, \bm{v}_{g_1}, \cdots, \bm{v}_{g_K}$ and $\bm{v}_{z}$, respectively. In order to combine these text representations and extract relationships among these representations, we use another Bi-LSTM to calculate a final text representation $\bm{v}$. To capture the visual features, for each physical object $x$ we integrate all its format information, including font family, font size, font color, bold, italic, centering and indent, into a vector $\bm{u}_x$. Thus, for the nodes of $c, g_1, \cdots, g_K$ and $z$, we get $\bm{u}_c, \bm{u}_{g_1}, \cdots, \bm{u}_{g_K}$ and $\bm{u}_z$, respectively.
Similarly, we use the third Bi-LSTM to get a final visual representation $\bm{u}$. Next, we concatenate $\bm{v}$ and $\bm{u}$, and send the combination vector into a feed-forward networks to obtain a synthetic representation. Finally, we use a Sigmoid function to obtain the probability $\hat{P}(c|ctx(s))$.

Readers may suggest that the context be expanded to consider all its other ancestors besides the immediate parent. However, this expansion definitely increases computational complexity.

\subsection{Inference}\label{sec:3.4}

For a new document, we aim to find out the optimal possible position sequence, $s_1^*, s_2^*, \cdots, s_N^*$ via maximizing the joint probability of inserting every physical object into a proper position. We have

\begin{equation*}
    \begin{aligned}
        (s_1^*, s_2^*, \cdots, s_N^*) = \mathop{\arg\max}_{s_1, s_2, \cdots, s_N} \sum_{i=1}^{N} \log P \left( c_i \ | \ ctx(s_i) \right),
    \end{aligned}
\end{equation*}
where $s_i \in S_i$ and $S_i$ represents the possible positions to insert $c_i$.

Note that searching the optimal path, $s_1^*, s_2^*, \cdots, s_N^*$, has exponential complexity, which makes the optimal result hard to search. Beam search is traditionally adopted for sequence or tree generation~\cite{zhou2018tree,sutskever2014sequence}, which considers multiple cases simultaneously in each step. In detail, we set a small integer $bs$ as the beam size, which represents the number of candidate trees. At each step, we extend each candidate tree in the beam with the top $bs$ most probable insertion positions. Thus, we obtain $bs \times bs$ candidate trees and remain the $bs$ most probable candidate trees according to their joint probability. When all the physical objects are inserted, we select the final hierarchical tree with the highest joint probability. When we set $bs=1$, the inference is the greedy method. The experiments in Section~\ref{sec:exp} will show that in different settings on $bs$ the greedy method achieves the best tradeoff between effectiveness and efficiency.

\subsection{Design Variants in HELD}\label{sec:varients}

\subsubsection{Traversal Orders of the Insertion Positions}

In the inference process, when inserting $c_i$ we check all the possible insertion positions in a certain order. Specifically, we propose three traversal methods: traversal-all, root-to-leaf and leaf-to-root, and introduce them using the example in Fig.~\ref{fig:generation_process}.

First, for the traversal-all method, we inquire about each insertion position and find out the position with the highest probability. Next, for the root-to-leaf method, we inquire about each insertion position in the order from the root to the leaf node. Once we find a position with the probability of more than 0.5, we return it as the result.  Finally, the leaf-to-root method is the same as the root-to-leaf method except that the traversal order changes to from the leaf to the root node. For the example in Fig.~\ref{fig:generation_process}, the results from these three methods are $p_2$, $p_1$, $p_3$, respectively. This example is deliberately generated to show the difference among them.

Note that different traversal methods involve different numbers of checks on the insertion positions. The theoretical analysis in Subsection~\ref{sec:app_proof} shows that to reach the proper insertion position the leaf-to-root method and traversal-all method check the smallest and largest number of positions, respectively. The results in Section~\ref{sec:exp} further empirically validate this. Additionally, the experimental results also show that the traversal-all and root-to-leaf methods achieve similar accuracy. Hence, the root-to-leaf method is the best choice.

\begin{figure*}
    \centering
    \includegraphics[width=4.0in]{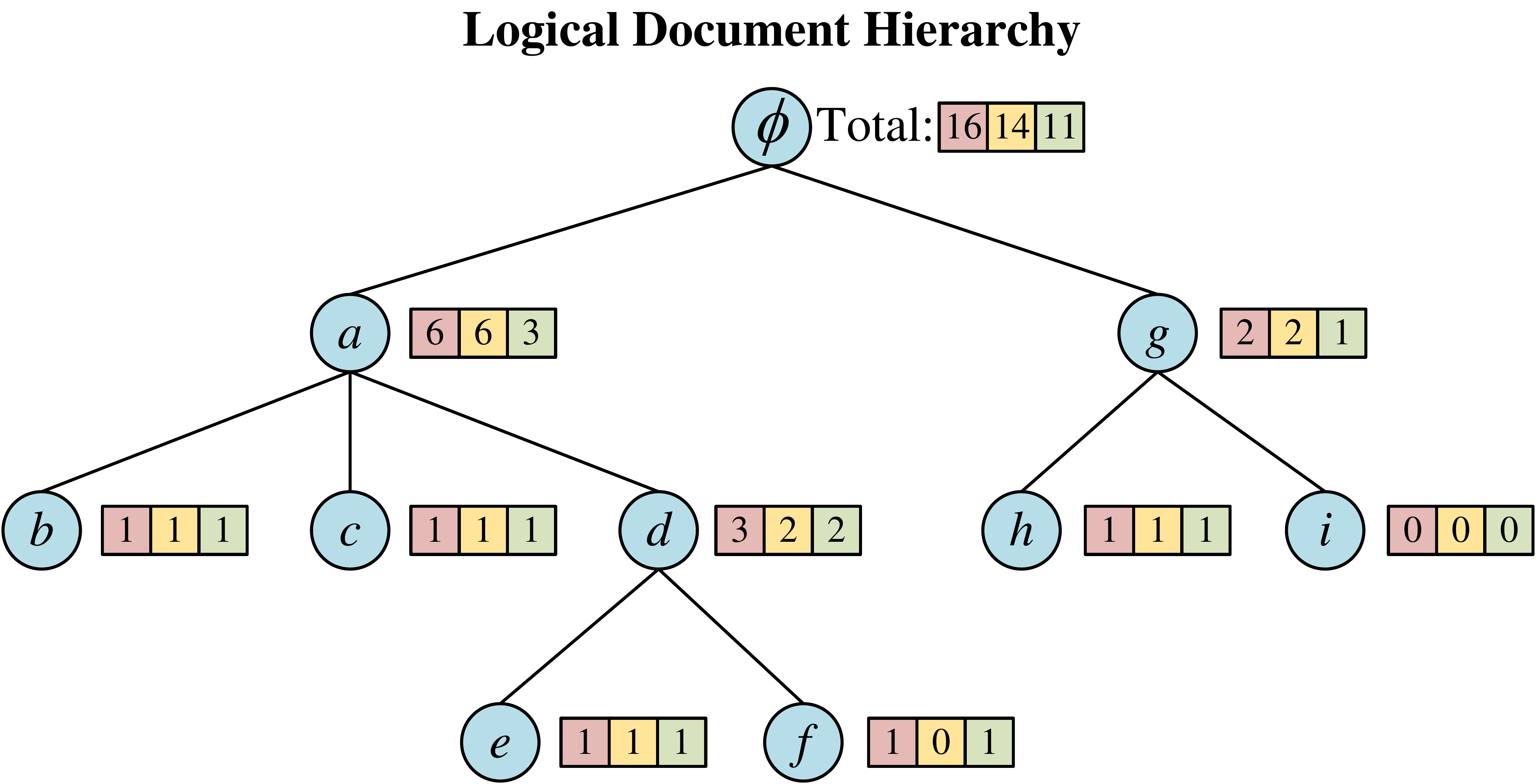}
    \caption{Inquiry number in hierarchy generation. The digits in the boxes represent total inquiry number of each node under different traversal methods. Specifically, red, yellow and green boxes represent the traversal-all, root-to-leaf and leaf-to-root method, respectively.}\label{fig:model_generation}
\end{figure*}

\subsubsection{Heading Extraction Explicitly or Implicitly}\label{sec:two_step}

We observe that all the internal nodes in the hierarchical tree $\mathcal{T}$ correspond to the headings of logical sections and all the leaf nodes of $\mathcal{T}$ correspond to concrete objects (e.g. paragraphs, tables, charts). Additionally, a section is usually semantically summarizable and visually observable by its heading~\cite{summers1998automatic}. Thus, another possible solution to our task is: first classify each physical object into heading or non-heading, second generate hierarchical tree for all the heading nodes, finally insert each non-heading object as the leaf child of its first previous heading object in the input sequence of physical objects. In other words, this new solution suggests a separated step of explicit heading extraction before the hierarchical tree generation.

We find that a separated step of explicit heading extraction brings about some benefits to our task. First, it might alleviate the difficulty of classification in the put-or-skip module, seeing the example in Fig.~\ref{fig:intro_diff_a} without a separated step of explicit heading extraction. Consider the insertion of node $g$, which is a heading. This node should be located at the same level as the node $f$, which is a non-heading object. For this situation that heading nodes and non-heading nodes are siblings, the model usually fails since heading and non-heading nodes usually have features of great differences. With an additional step of heading extraction, tree building is much easier for only the heading nodes. The experiments in Section~\ref{sec:exp} further validate that this two-step solution increases the model accuracy.

Second, although heading extraction introduces additional computing overhead the following computing of node insertion will greatly decrease. Specifically, the model is applied to the heading nodes for tree generation, and the other non-heading nodes are inserted by the rule. Overall, this two-step solution gains much increase in time efficiency, which is also demonstrated by the experiments.

Note that distinguishing heading nodes from non-heading ones can be formulated as a sequence labeling task. In detail, we use Bi-LSTM~\cite{schuster1997bidirectional} to extract textual and format features of the local context and then apply multi-layers CNNs~\cite{simonyan2014very} to consider the long-distance association. Inspired by the previous work~\cite{tan2018deep}, we add a self-attention layer~\cite{vaswani2017attention}. Finally, a Sigmoid layer is used to classify whether a physical object is heading or not.

\subsubsection{Tolerance to Insertion Errors in Predecessor Steps}\label{sec:tolerance}

Note that in the actual inference for a document, before inserting a node $c_i$ some previous nodes in $\{c_1, , \cdots, c_{i-1}\}$ might be inserted into wrong positions. Thus, we need to deliberately make some training data with such insertion errors. Specifically, we simulate the tree-building process with some random insertion errors, where a node is inserted into one of the other possible positions except the right one. Based on the resultant tree, some extra training data can be generated accordingly. The experimental results show that these new training data bring about significant improvements in terms of effectiveness.

\begin{figure*}[!htb]
    \centering
    \subfigure[]{\label{fig:level_number_zh}
    \includegraphics[width=4.0in]{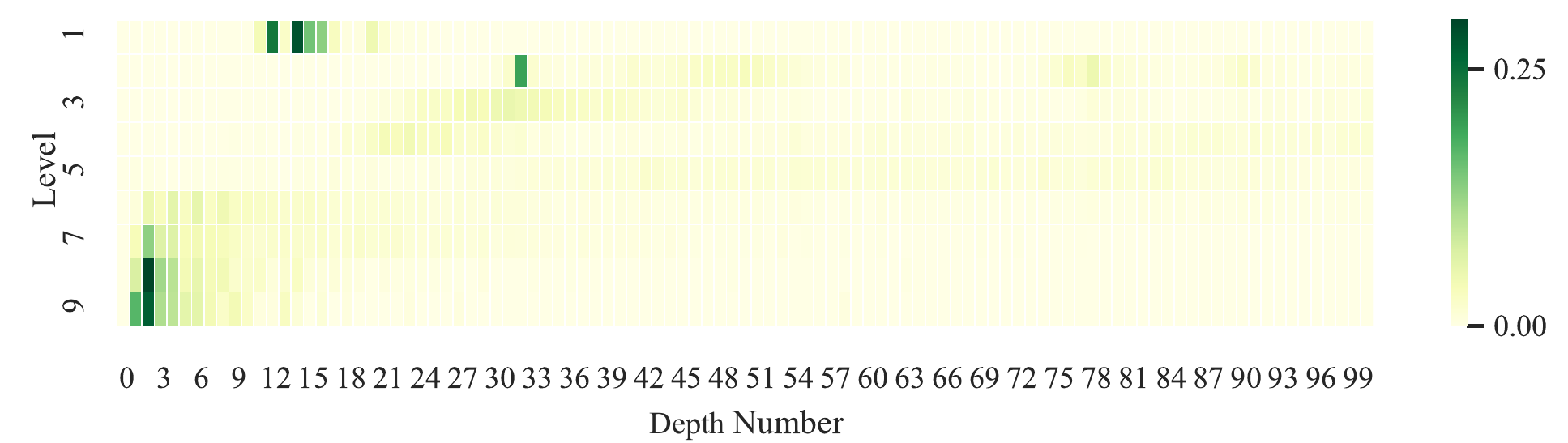}}
    \subfigure[]{\label{fig:level_number_en}
    \includegraphics[width=4.0in]{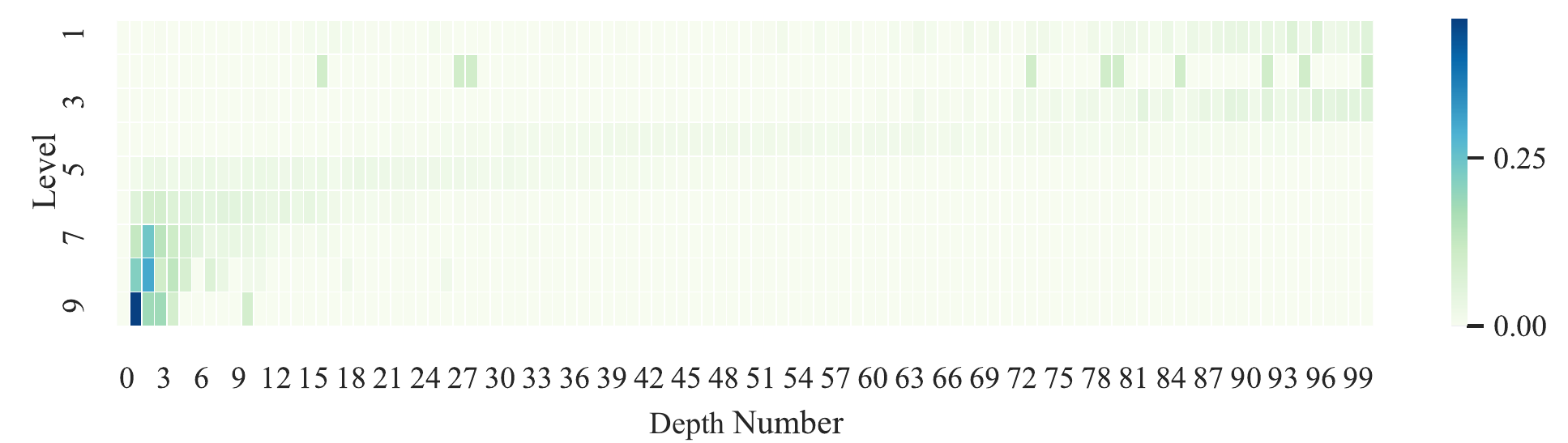}}
    \subfigure[]{\label{fig:level_number_arxiv}
    \includegraphics[width=4.0in]{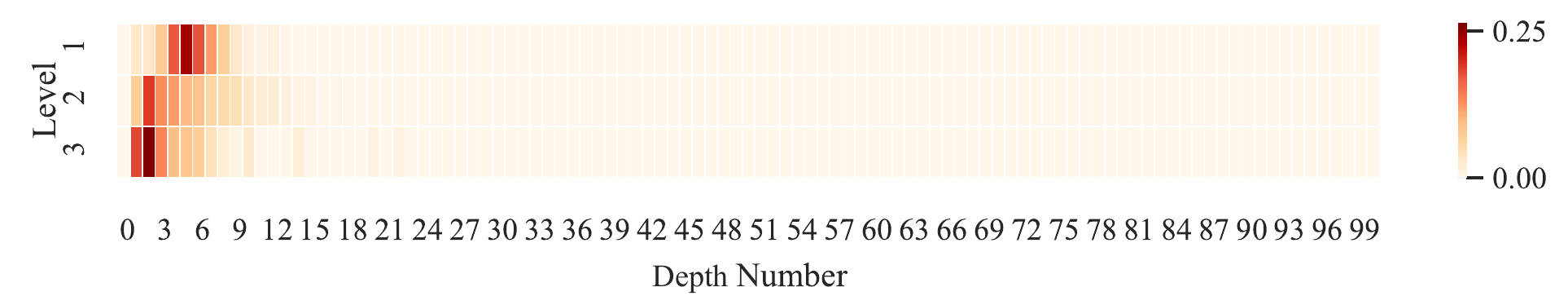}}
    \caption{We calculate the distribution of the number of nodes at different depth in the Chinese, English and arXiv datasets. (a) Depth-number distribution on Chinese dataset. (b) Depth-number distribution on English dataset. (c) Depth-number distribution on arXiv dataset.}
    \label{fig:heading_observation_a}
\end{figure*}

\begin{figure*}[!htb]
    \centering
    \subfigure[]{\label{fig:level_pattern_zh}
    \includegraphics[width=4.0in]{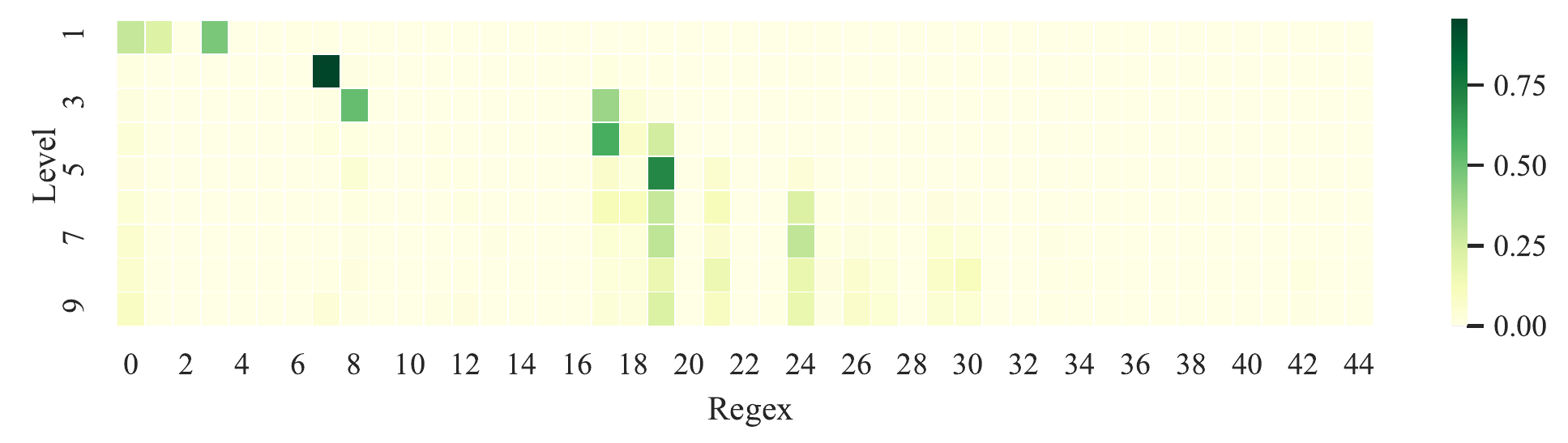}}
    \subfigure[]{\label{fig:level_pattern_en}
    \includegraphics[width=4.0in]{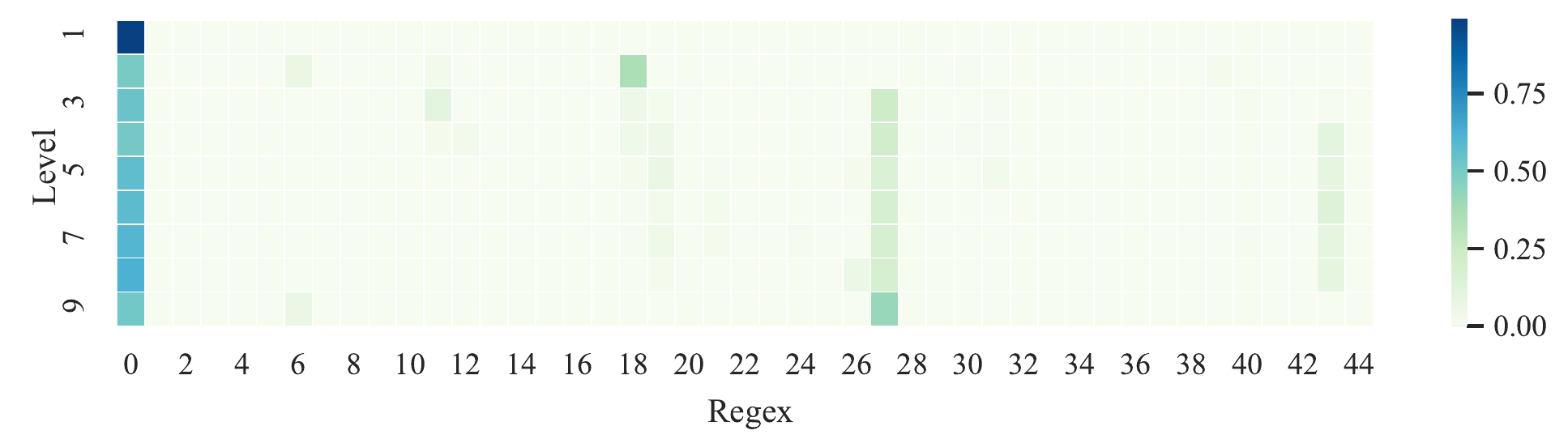}}
    \subfigure[]{\label{fig:level_pattern_arxiv}
    \includegraphics[width=4.0in]{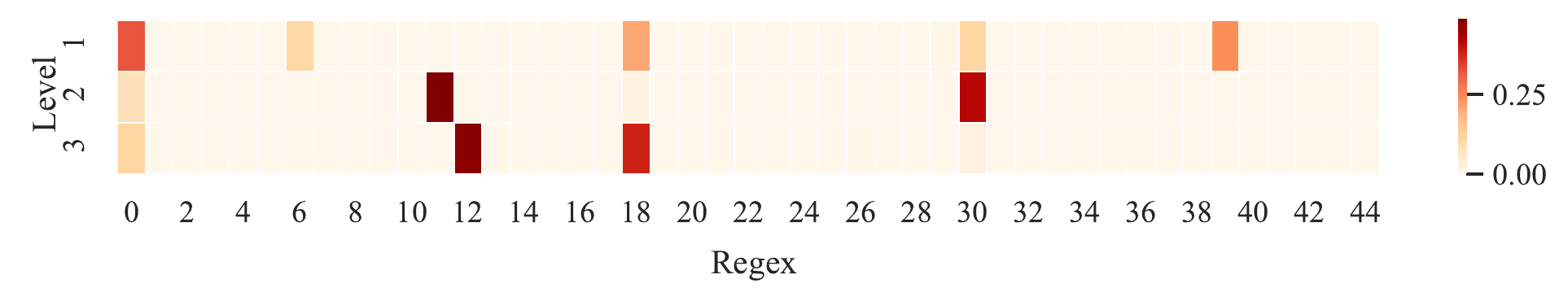}}
    \caption{We calculate the distribution of the number of matched key word-level features at different depth in the Chinese, English and arXiv datasets. (a) Depth-regex distribution on Chinese dataset. (b) Depth-regex distribution on English dataset. (c) Depth-regex distribution on arXiv dataset.}
    \label{fig:heading_observation_b}
\end{figure*}

\subsection{Theoretical analysis on the efficiency of different traversal methods}\label{sec:app_proof}

In the following, for each traversal method, we will theoretically count the number of checks on the insertion positions, which is required to reach the ground-truth position. Note that in this analysis we assume that the ground-truth position for inserting each node is provided. This number is equal to the one when an ideal model with 100\% accuracy is adopted in inference.

Specifically, after estimating $P(c|ctx(s))$ the parent node of $s$ will be counted once. To sum up the numbers on all the nodes, we get the final total, denoted as $N_{\text{all}}, N_{\text{r2l}}$ and $N_{\text{l2r}}$ for the three methods of traversal-all, root-to-leaf and leaf-to-root. We have

{\small
\begin{flalign*}
    N_{\text{all}} &= \sum_{i=1}^{N} (s_i + \mathbbm{1}(\text{node } i \text{ is not on the rightmost-branch}) ), \\
    N_{\text{r2l}} &= \sum_{i=1}^{N} (s_i + \mathbbm{1}(\text{node } i \text{ has next sibling})), \\
    N_{\text{l2r}} &= \sum_{i=1}^{N} (s'_i + \mathbbm{1}(\text{node } i \text{ is not on the rightmost-branch})),
\end{flalign*}}where $s_i$ and $s_i'$ refer to the number of all the descendants and non-leaf descendants of a node $i$ in the tree, $N$ refers to the number of nodes in the tree, $\mathbbm{1}(\cdot)$ is the indicator function. Fig.~\ref{fig:model_generation} shows an example with the inquiry numbers for these three methods.

Next, we calculate the relationships between these numbers as follow,

{\small
\begin{flalign*}
    N_{\text{all}} - N_{\text{r2l}} &= I - l \gg 0, \\
    N_{\text{all}} - N_{\text{l2r}} &= \sum_{i=1}^{N} (s_i - s'_i) \gg L \gg I > N_{\text{all}} - N_{\text{r2l}},
\end{flalign*}}where $I$ and $L$ is the number of the internal node and leaf node in the tree respectively, $l$ is the length of its rightmost-branch.

Usually, we have $L \gg I \gg l$. This is also true for all the hierarchical trees used in this study. Thus, we have $N_{\text{l2r}} < N_{\text{r2l}} < N_{\text{all}}$.

Finally, we argue that although these three numbers might not equal the inquiry numbers with the actual model when its accuracy is less than 100\%, they are a good approximation of the actual number. This is also demonstrated in the experiments in Section~\ref{sec:exp}. 

\section{Experiment Details}

\subsection{Dataset}

Since the documents in published datasets for our task~\cite{Pembe2010,Manabe2015,Rahman2017Understanding,Bentabet2019} only contain tens of pages and have shallow (4 levels at the most) logical hierarchy, we build three datasets with variable-depth logical hierarchy from long documents: 1) the Chinese dataset that contains prospectuses and annual reports from the China Securities Exchange market, 2) the English dataset that contains annual reports from Hongkong Exchange market, and 3) the arXiv dataset that contains English scientific publications from arXiv. The documents in the Chinese and English dataset can be downloaded from CNINFO\footnote{\url{http://www.cninfo.com.cn/}. Last visited in 2021/04/28.} and the documents in the arXiv dataset can be downloaded from arXiv\footnote{\url{https://arxiv.org/}. Last visited in 2021/04/28.}.

Each document is assigned to at least two annotators for annotating its logical document hierarchy.
If the results on a document are different, another senior annotator will address the conflicts and output the final
answer.

For three datasets, we split training, validation and test set with around 8:1:1 split.
For the Chinese dataset, we split 1030 documents into 830 for training, 100 for validation and 100 for test.
For the English dataset, we split 1110 documents into 910 for training, 100 for validation and 100 for test.
For the arXiv dataset, we split 1732 documents into 1432 for training, 150 for validation and 150 for test.

\subsection{Dataset Analysis}

\begin{figure*}
    \centering
    \subfigure[]{
    \includegraphics[width=1.45in]{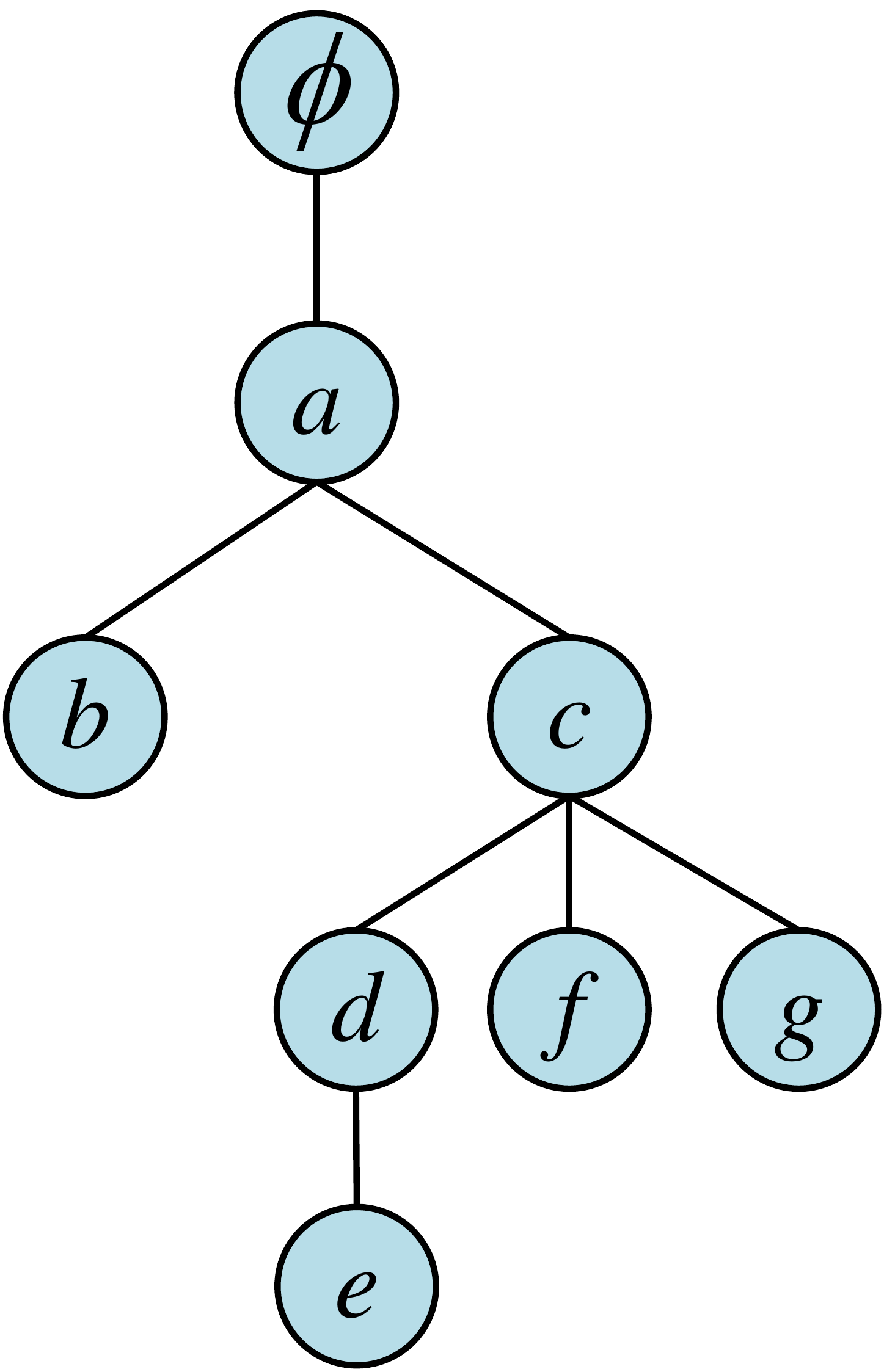}}
    \subfigure[]{
    \includegraphics[width=1.7in]{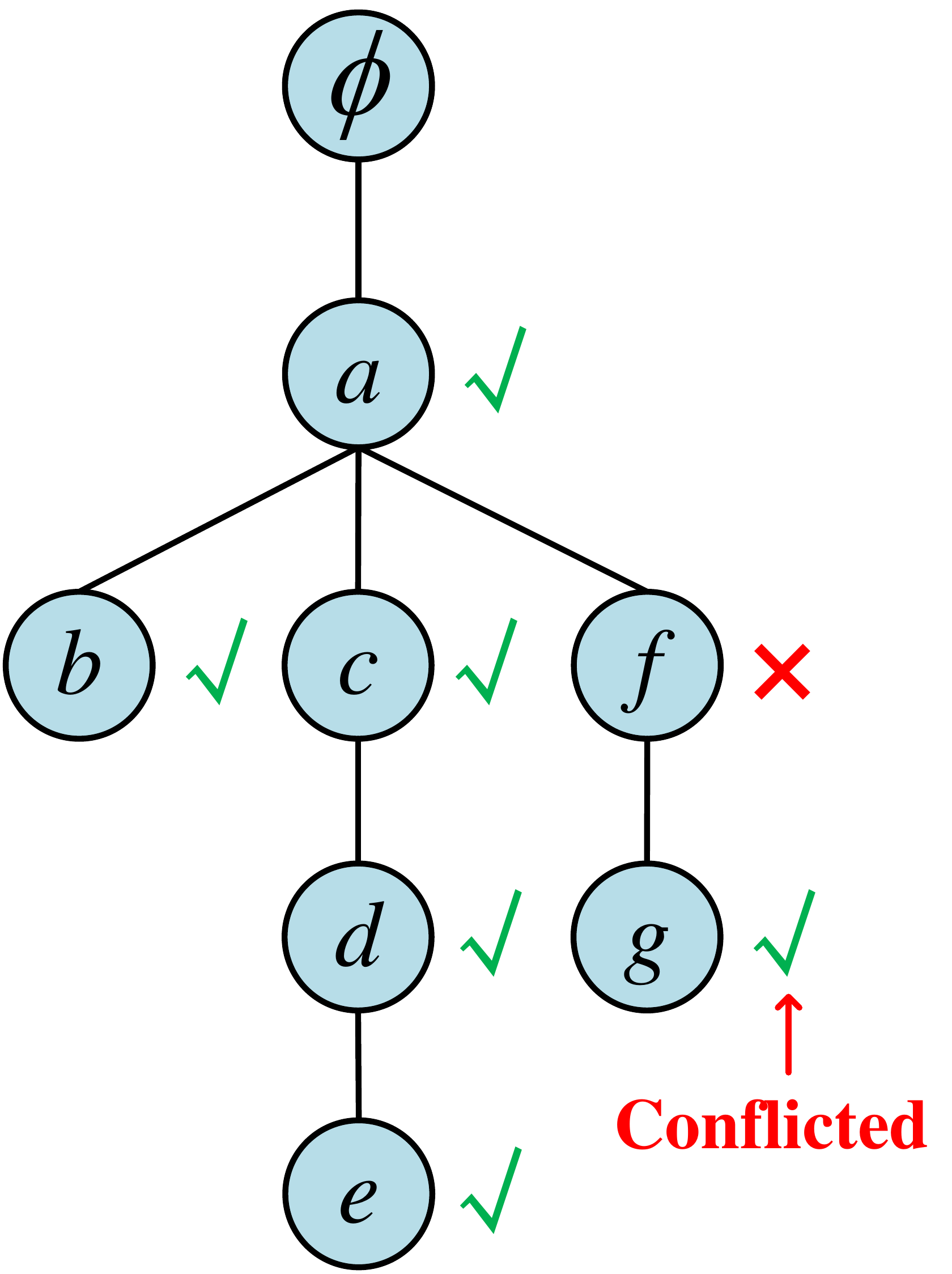}}
    \subfigure[]{
    \includegraphics[width=1.5in]{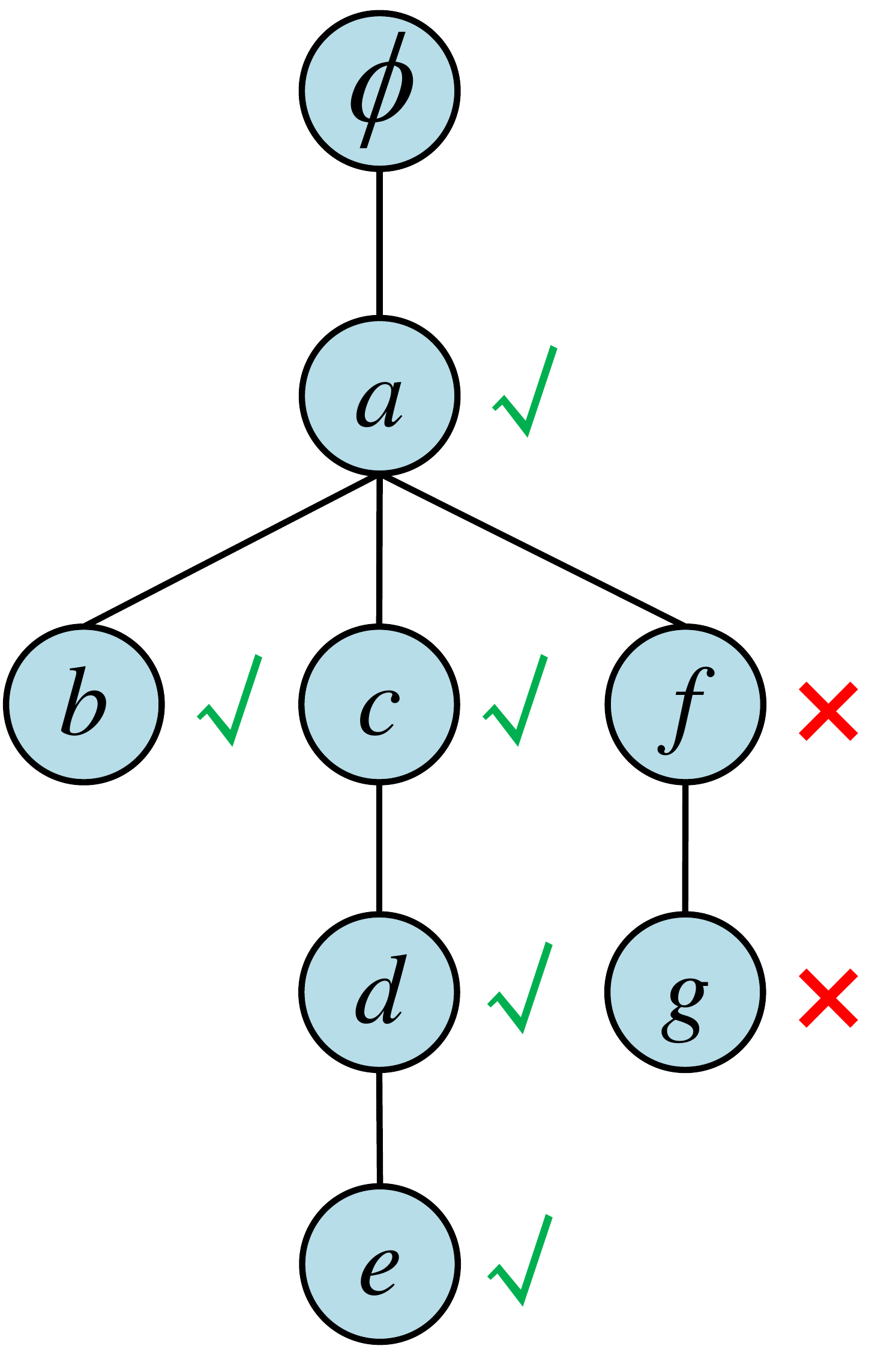}}
    \captionof{figure}{The comparison of old and new evaluation metrics. (a) The ground-truth hierarchy. (b) The predicted hierarchy and the correctness of each node under previous metric. (c) The predicted hierarchy and the correctness of each node under our metric.}\label{fig:evaluation}
\end{figure*}

In this subsection, we further analyze these three datasets and illustrate some observations in Fig.~\ref{fig:heading_observation_a} and Fig.~\ref{fig:heading_observation_b}.

On the one hand, we calculate the number of internal nodes at each level in each document and depicit the distribution in Fig.~\ref{fig:level_number_zh}, Fig.~\ref{fig:level_number_en} and Fig.~\ref{fig:level_number_arxiv} for three datasets, respectively.
For example, in the Chinese dataset, we find that most documents have 10 - 20 headings in level 1.
However, for levels 3 to 7, most documents have a different number of headings, therefore the distribution is very uniform.
In the English dataset, for every level, most documents have a different number of headings, which means that the degree of difference in the English dataset is greater than that in the Chinese dataset.
In the arXiv dataset, the number of headings at each depth is relatively concentrated, since the logical hierarchies in scientific publications are usually pre-defined.

On the other hand, we aim to observe whether these documents follow some types of templates.
First, we conclude 44 types of patterns to represent item number in headings.
For example, headings $c, e, m, o$ in Fig.~\ref{fig:intro_diff_a} follow the pattern of ``(1), (2), $\cdots$''.
Headings $b, h, i, l$ in Fig.~\ref{fig:intro_diff_a} follow the pattern of ``1., 2., $\cdots$''.
For each type of pattern, we can design one regex to match it, thus, we design 44 regexes to represent these patterns.
Then, for a given document hierarchy, we can check each internal node in this hierarchy and decide whether it matches one of the 44 regexes.
Thus, for the internal nodes on each level in all the documents in the dataset, we count the matching ratio of these 44 regexes and draw the distribution graph in Fig.~\ref{fig:level_pattern_zh}, Fig.~\ref{fig:level_pattern_en} and Fig.~\ref{fig:level_pattern_arxiv}.
For example in Fig.~\ref{fig:level_pattern_zh}, we observe that most internal nodes in the first level match regexes 0, 1 and 3. Here, regex 0 means the node matches none of the 44 regexes. Almost all the internal nodes in the second level match regex 7.
Clearly, in some deeper levels, such as levels 5 - 9, these internal nodes match more various regexes.
Especially, in the English dataset, most headings do not match any regexes.
Therefore, the hierarchies in these documents have differences to some degree, therefore the hierarchy is not definitely pre-defined.
Moreover, the degree of difference in the English dataset is greater than that in the Chinese dataset.

\subsection{Evaluation Methods}\label{sec:evaluation}

In this study, we propose a new metric to judge whether a certain node is inserted correctly or not. First, we define that physical object $c_i$ is inserted correctly if its predicted path $\hat{r}_i$ completely equals to its ground-truth path $r_i$. Here, the path of $c_i$ means an ordered node list that consists of $c_i$ and all its ancestors up to the root $\phi$. Note that previous studies~\cite{Luong2010Logical,Constantin2013PDFX,Tkaczyk2015CERMINE} define that a node $c_i$ is inserted correctly if it is put at the right level of the tree. They ignore the error that the depth of a node is correct while its path to the root is wrong. For example, in Fig.~\ref{fig:evaluation} node $g$ is considered as an error by the new measure since its path to the root, namely $(g\rightarrow f \rightarrow  a \rightarrow \phi)$, is not equal to the ground-truth path $(g\rightarrow c \rightarrow  a \rightarrow \phi)$. However, it is considered as a correct insertion since it is put into the right level of 3.

With this new definition of correct insertion, we can calculate the overall accuracy of node insertion. Additionally, for any tree level $k$ we can also calculate the precision, recall, and F1 denoted as $p_k, r_k$ and $F1_k$, respectively. In the experimental results, we use these measures in micro average for all the testing documents.

To evaluate the efficiency in processing new documents, we record the average execution time of each document. Additionally, we also count the number of calls for the put-or-skip module, denoted as "\#inquiry", since they are the major time consumption.

\subsection{Baselines}\label{sec:detail}

\begin{table*}[ht!]
    \centering
    \caption{Comparing HELD and baseline models on the test set of the Chinese dataset.}
    \label{table:baseline_chn}
    \scriptsize
    \begin{tabular}{lcccccccccccc}
        \hline
        Model & $acc$ & $F1_1$ & $F1_2$ & $F1_3$ & $F1_4$ & $F1_5$ & $F1_6$ & $F1_7$ & $F1_8$ & $F1_9$ & $F1_{10}$ & $F1_{11}$ \\
        \hline
        HEPS~\cite{Manabe2015} & 0.3764 & 0.8030 & 0.6794 & 0.4913 & 0.4781 & 0.2926 & 0.1287 & 0.0138 & 0.0000 & 0.0000 & 0.0000 & 0.0000 \\
        TOC~\cite{Bentabet2019} & 0.9403 & 0.9804 & 0.9408 & 0.9562 & 0.9632 & 0.9528 & 0.9333 & 0.8512 & 0.6305 & 0.2645 & 0.0000 & 0.0000 \\
        Pembe's~\cite{Pembe2010} & 0.9339 & 0.9573 & 0.9445 & 0.9674 & 0.9666 & 0.9501 & 0.9305 & 0.8627 & 0.6944 & 0.4418 & 0.0629 & 0.0000 \\
        1step-HELD & 0.9583 & 0.9918 & 0.9577 & 0.9748 & 0.9695 & 0.9617 & 0.9486 & 0.9179 & 0.8693 & 0.6810 & 0.4620 & 0.2483 \\
        2step-HELD (l2r) & 0.9720 & 0.9892 & 0.9483 & 0.9769 & 0.9831 & 0.9761 & 0.9662 & 0.9471 & 0.9145 & 0.8299 & 0.8491 & 0.7059 \\
        2step-HELD (r2l) & 0.9726 & 0.9892 & 0.9486 & 0.9779 & 0.9832 & 0.9771 & 0.9670 & 0.9454 & 0.9179 & 0.8531 & 0.7950 & 0.7347 \\
        2step-HELD (ta) & 0.9731 & 0.9892 & 0.9489 & 0.9784 & 0.9838 & 0.9778 & 0.9675 & 0.9449 & 0.9174 & 0.8521 & 0.8517 & 0.7059 \\
        - Tolerance Errors & 0.9725 & 0.9892 & 0.9481 & 0.9769 & 0.9833 & 0.9768 & 0.9671 & 0.9481 & 0.9138 & 0.8501 & 0.8517 & 0.7059 \\
        \hline
    \end{tabular}
\end{table*}

\begin{table*}[ht!]
    \centering
    \caption{Comparing HELD and baseline models on the test set of the English dataset.}
    \label{table:baseline_eng}
    \scriptsize
    \begin{tabular}{lcccccccccc}
        \hline
        Model & $acc$ & $F1_1$ & $F1_2$ & $F1_3$ & $F1_4$ & $F1_5$ & $F1_6$ & $F1_7$ & $F1_8$ & $F1_9$ \\
        \hline
        HEPS~\cite{Manabe2015} & 0.4779 & 0.6882 & 0.6630 & 0.6265 & 0.4871 & 0.2976 & 0.1744 & 0.0525 & 0.0187 & 0.0000 \\
        TOC~\cite{Bentabet2019} & 0.6436 & 0.8296 & 0.7928 & 0.7785 & 0.6190 & 0.3671 & 0.2111 & 0.0776 & 0.0000 & 0.0000 \\
        Pembe's~\cite{Pembe2010} & 0.6563 & 0.8062 & 0.7964 & 0.7716 & 0.6460 & 0.4364 & 0.3286 & 0.2337 & 0.0602 & 0.0000 \\
        1step-HELD & 0.6117 & 0.7008 & 0.6865 & 0.7053 & 0.5984 & 0.4830 & 0.3591 & 0.2600 & 0.0485 & 0.0000 \\
        2step-HELD (l2r) & 0.7075 & 0.8555 & 0.8365 & 0.8119 & 0.7026 & 0.5577 & 0.4200 & 0.3136 & 0.3296 & 0.3982 \\
        2step-HELD (r2l) & 0.7291 & 0.8264 & 0.8250 & 0.8133 & 0.7196 & 0.6022 & 0.4658 & 0.4738 & 0.4883 & 0.5325 \\
        2step-HELD (ta) & 0.7301 & 0.8556 & 0.8380 & 0.8180 & 0.7211 & 0.6048 & 0.4807 & 0.3763 & 0.4258 & 0.4245 \\
        - Tolerance Errors & 0.7095 & 0.8324 & 0.8120 & 0.7942 & 0.6955 & 0.5867 & 0.4646 & 0.4469 & 0.3686 & 0.3719 \\
        \hline
    \end{tabular}
\end{table*}

\begin{table*}[ht!]
    \centering
    \caption{Comparing HELD and baseline models on the test set of the arXiv dataset.}
    \label{table:baseline_arxiv}
    \scriptsize
    \begin{tabular}{lccccc}
        \hline
        Model & $acc$ & $F1_1$ & $F1_2$ & $F1_3$ & $F1_4$ \\
        \hline
        HEPS~\cite{Manabe2015} & 0.8375 & 0.9385 & 0.8975 & 0.5963 & 0.3218 \\
        TOC~\cite{Bentabet2019} & 0.8908 & 0.9800 & 0.9301 & 0.7634 & 0.5828 \\
        Pembe's~\cite{Pembe2010} & 0.9034 & 0.9734 & 0.9243 & 0.8236 & 0.6842 \\
        2step-HELD (l2r) & 0.9546 & 0.9923 & 0.9705 & 0.9032 & 0.7069 \\
        2step-HELD (r2l) & 0.9567 & 0.9926 & 0.9720 & 0.9073 & 0.7069 \\
        2step-HELD (ta) & 0.9578 & 0.9926 & 0.9730 & 0.9104 & 0.7178 \\
        \hline
    \end{tabular}
\end{table*}

\paragraph{HEPS Model~\cite{Manabe2015}} This is a rule-based method for our task. The original HEPS model focuses on web pages. Some of the features from web pages cannot be obtained in PDF files. Thus, this method is tailored to use only the features in PDF files.

\paragraph{TOC Model~\cite{Bentabet2019}}
TOC model~\cite{Bentabet2019} is a sequence labeling-based model. Since the document length is greatly longer than the documents in the TOC model, we use CNNs to replace LSTMs in sequence labeling to improve efficiency.

\paragraph{Pembe's Model~\cite{Pembe2010}}
Pembe's model is a tree generation-based model. In this method logistic regression is adopted to select the proper position for node insertions.

\subsection{Hyper-parameters configuration}

Here, we introduce some hyper-parameters of the proposed HELD model. First, we use skip-gram~\cite{Mikolov2013Efficient} to pre-train character embeddings with 24-dimension. In heading recognition step, we use a 9-layer Network In the Network~\cite{lin2013network} to extract contextual features. The kernel size of each layer is 5, 1, 5, 5, 1, 5, 5, 1, 5. The kernel number of each layer is 128, 64, 128, 256, 128, 256, 512, 256, 512. We adopt batch normalization~\cite{ioffe2015batch} immediately after each convolution and before all ReLU activate functions~\cite{nair2010rectified}. In put-or-skip model (according to Subsection~\ref{sec:3.3}), we set hidden dimension as 128, 512 and 64 for $v_x$, $v_T$ and $u_F$, respectively. We use ``tanh'' activate function in LSTMs. We set hidden dimension as 128 for FNN layer. Then, weight initialization in~\cite{he2015delving} is used to initialize parameters and Adam~\cite{kingma2014adam} optimizer is used to update parameters. We set mini-batch size as $128$ and learning rate as $0.00005$. We train the model on 2 Titan 1080Ti GPUs and use Horovod~\cite{sergeev2018horovod} to update parameters in a distributed way. 

\section{Experiment Results}
\label{sec:exp}

\subsection{Results}

In this subsection, we aim to answer these research questions:

$\bullet$ RQ1: What is the effectiveness of HELD model compared with other baselines?

$\bullet$ RQ2: What is the effectiveness and efficiency of different traversal methods in HELD model?

$\bullet$ RQ3: What are the effectiveness and efficiency of the one-step and two-step framework?

$\bullet$ RQ4: What is the effectiveness of adding tolerance to insertion errors in predecessor steps?

$\bullet$ RQ5: What is the effectiveness of different features in the put-or-skip module?

$\bullet$ RQ6: What is the effectiveness and efficiency of beam size?

$\bullet$ RQ7: What is the influence of noise in document layout recognition?

\textbf{For RQ1}, we compare the proposed two-step HELD model with three baseline models, by evaluating $F1_k$ on the test set of three datasets. The results are shown in Table~\ref{table:baseline_chn}, Table~\ref{table:baseline_eng} and Table~\ref{table:baseline_arxiv}, respectively.

Note that, we extract headings explicitly (two-step), use the traversal-all method and set beam size as 1 in the inference process to obtain the best HELD model. The proposed HELD obtains 0.9731, 0.9301 and 0.9578 node accuracy on each dataset, respectively. However, the HEPS model obtains 0.3764, 0.4779, 0.8375 node accuracy, the TOC model obtains 0.9403, 0.6436, 0.8905 node accuracy and the Pembe's model obtains 0.9339, 0.6563, 0.9034 node accuracy, respectively. Clearly, on each dataset, the proposed HELD model has great improvement on the F1 value of every level ($F1_k$) and total node accuracy ($acc$) compared with three baseline models.
Note that, compared with the other two datasets, all the models obtain lower accuracy on the English dataset, since the visual and textual cues are more implicit in this dataset. Like the intuitive analysis in Section~\ref{sec:intro}, the rule-based model (HEPS model) obtains low accuracy since the assumptions in this model are not always true.
Sequence labeling-based model (TOC model) cannot predict well for the physical objects on deep levels since it considers hierarchical depth as an absolute concept and neglects the containment and parallel relation between physical objects.
As shown in Fig.~\ref{fig:level_pattern_zh}, Fig.~\ref{fig:level_pattern_en} and Fig.~\ref{fig:level_pattern_arxiv}, headings at the same level match different regexes and headings that match the same regex locate at different levels.
That is to say, it is hard to directly predict the level of each heading, especially for the headings on deeper levels.
Therefore, the TOC model obtains lower accuracy, especially for the headings on deeper levels.
Pembe's model is based on hierarchy generation, however, it is not good at extracting format and semantic features, thus it underperforms the proposed HELD model.
Especially on the English dataset, many headings match no regex as shown in Fig.~\ref{fig:level_pattern_en}, thus predicting the level of these headings depends on both format and semantic features.
Therefore, Pembe's model obtains low accuracy on the English dataset.
By combining format and semantic features to extract containment and parallel relation between physical objects, the proposed HELD model outperforms the other baselines and obtains the node accuracy of 0.9731 and 0.7301 on Chinese and English datasets, respectively.

\begin{figure*}[!htb]
    \centering
    \subfigure[]{\label{fig:traversal_comp_efficiency}
    \includegraphics[width=3.0in]{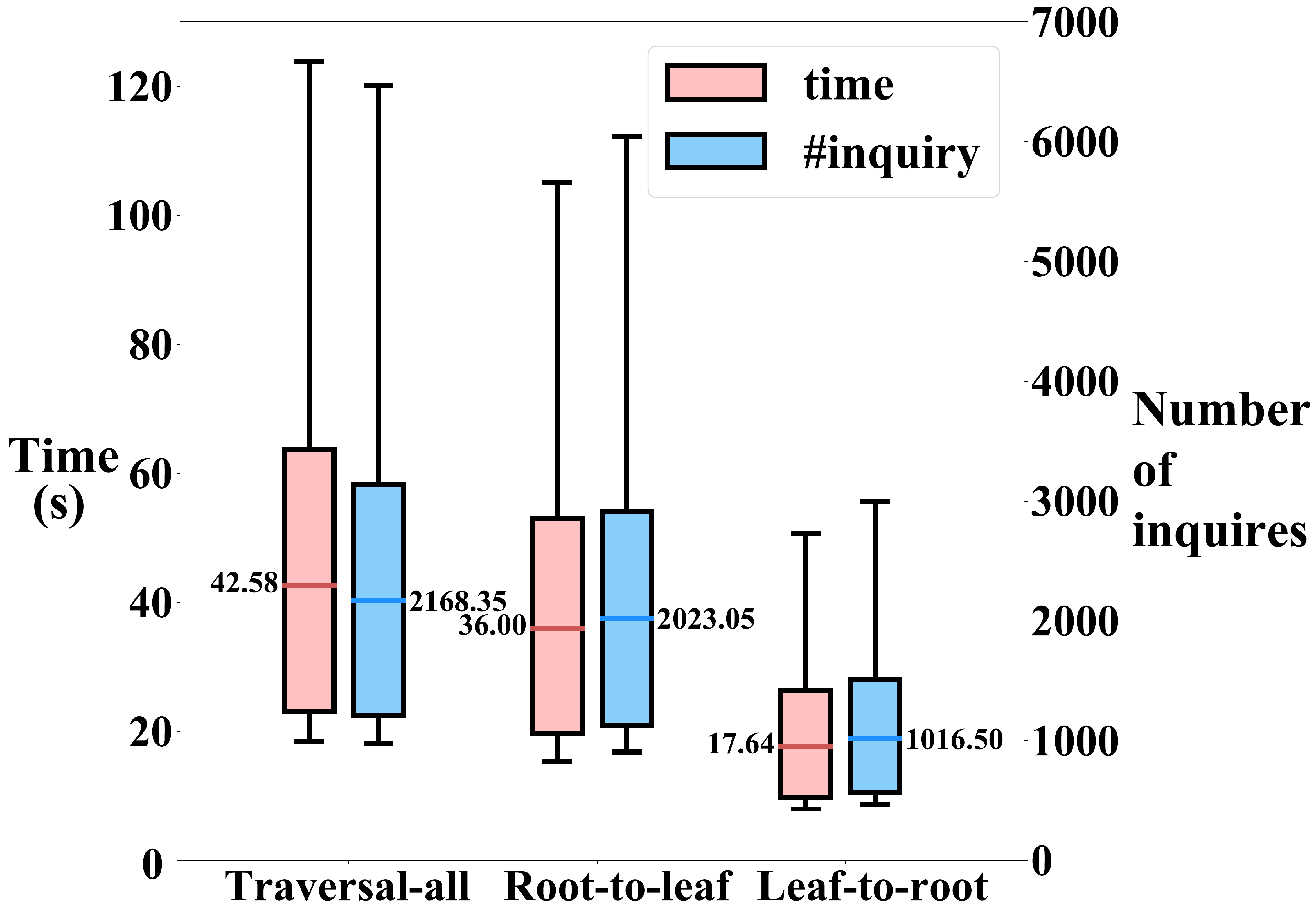}}
    \subfigure[]{\label{fig:model_comp_efficiency}
    \includegraphics[width=3.0in]{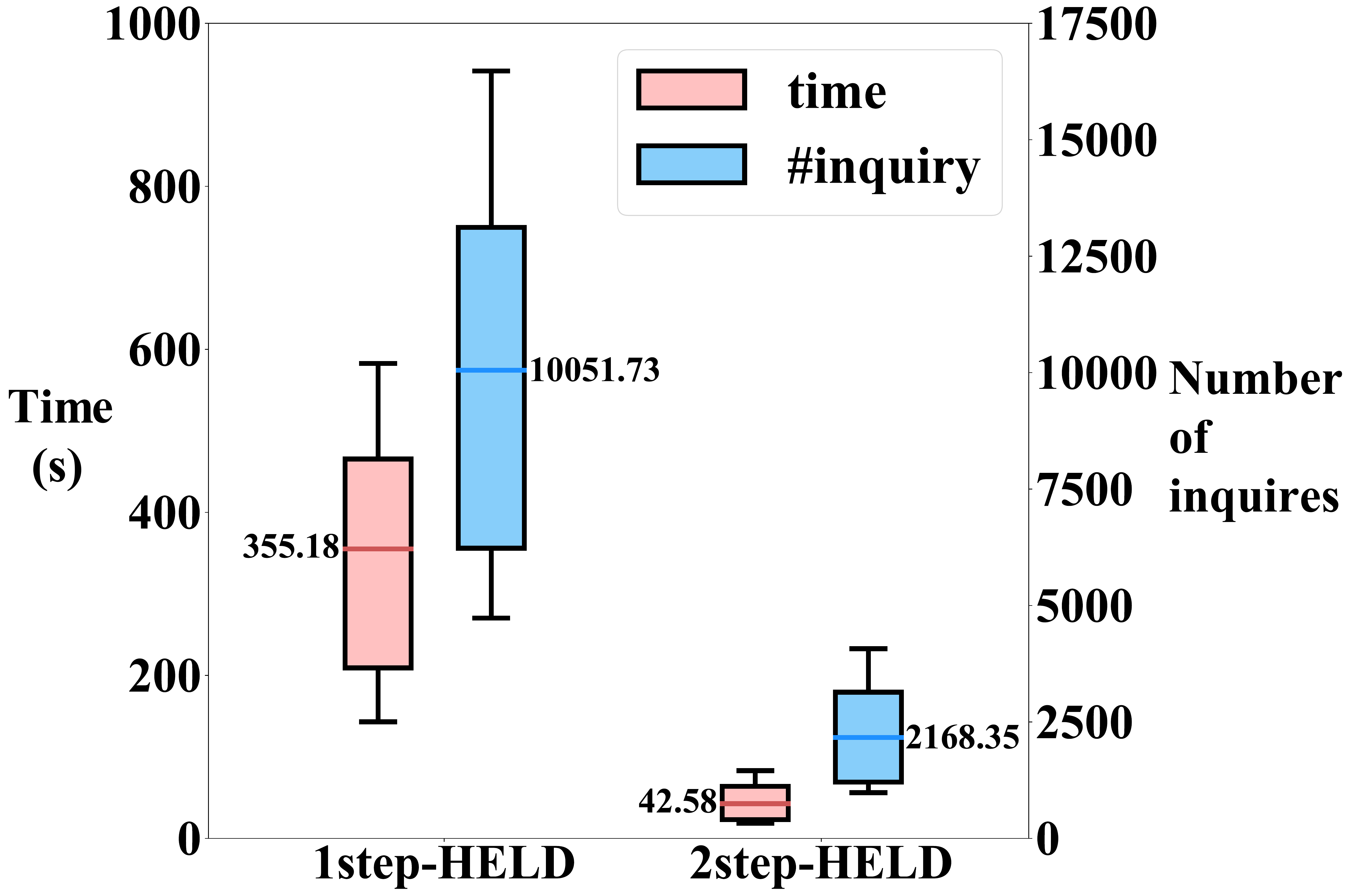}}
    \caption{Results on efficiency of different models. (a) Comparing different traversal methods in HELD model. (b) Comparing 1step-HELD model and 2step-HELD model.}
    \label{fig:model_comp}
\end{figure*}

\textbf{For RQ2}, we compare three different traversal methods in the HELD model, by evaluating $F1_k$ and \#inquiry on the test set of three datasets. The results are shown in Fig.~\ref{fig:traversal_comp_efficiency}, Table~\ref{table:baseline_chn}, Table~\ref{table:baseline_eng} and Table~\ref{table:baseline_arxiv}, respectively.

The traversal-all method obtains accuracy of 0.9731, 0.7301 and 0.9578 on Chinese, English and arXiv datasets, respectively. Root-to-leaf method obtains accuracy of 0.9726, 0.7291 and 0.9567 on Chinese, English and arXiv datasets, respectively.
Apparently, the traversal-all method outperforms the root-to-leaf method, since the traversal-all method inquiries all the possible insertion positions, however, the gap between them is subtle. Note that the leaf-to-root method obtains lower accuracy than other traversal methods. Because the root-to-leaf method assigns lower priority for those physical objects on the shallow level, these nodes on the shallow level have higher importance than the other nodes. The reason is that if the parent node is inserted into incorrect positions, all of its descendant nodes will be wrong under our evaluation measure.

To explore the efficiency of three traversal methods, we count the processing time for an average document and \#inquiry in Fig.~\ref{fig:traversal_comp_efficiency}. The traversal-all method consumes 42.58 seconds and 2168.35 \#inquiry on average to process a document. For comparison, the root-to-leaf method consumes 36.00 seconds and 2023.05 \#inquiry to process a document on average (obtaining 1.2x speedup ratio) and the leaf-to-root method only consumes 17.64 seconds and 1016.5 \#inquiry to process a document on average (obtaining 2.4x speedup ratio). Note that, the efficiency order of three traversal methods in practice is ``leaf-to-root $>$ root-to-leaf $>$ traversal-all'', which empirically validates the efficiency order in theory (shown in Subsection~\ref{sec:app_proof}).

Therefore, we can use the traversal-all method if higher accuracy is required, and use the root-to-leaf method if higher efficiency is required. Since the leaf-to-root method obtains great improvement on efficiency, it can be used if the requirement of efficiency is much higher than effectiveness.

\textbf{For RQ3}, we compare one-step HELD with two-step HELD, by evaluating $F1_k$ and \#inquiry on the test set of three datasets. The results are shown in Fig.~\ref{fig:model_comp_efficiency}, Table~\ref{table:baseline_chn}, Table~\ref{table:baseline_eng} and Table~\ref{table:baseline_arxiv}, respectively.

Apparently, the best 2step-HELD model obtains node accuracy of 0.9731 and 0.7301 on the Chinese and English datasets, respectively. By comparison, the 1step-HELD model obtains node accuracy of 0.9583 and 0.6117 on the Chinese and English datasets, respectively. In other words, the 2step-HELD model greatly outperforms the 1step-HELD model on effectiveness. The reason is that generating hierarchy based on heading objects sequence alleviates the difficulty of classification in the put-or-skip module (according to analysis in Subsection~\ref{sec:two_step}). Note that, the 1step-HELD model obtains higher $F1_1$ and $F1_2$ than the 2step-HELD model on the Chinese dataset, because the 2step-HELD model might make some classification mistakes in heading recognition step. In general, extracting heading objects explicitly obtains great improvement on effectiveness.

From the efficiency view, as shown in Fig.~\ref{fig:model_comp_efficiency}, the 2step-HELD model consumes less execution time and demands fewer inquiry number. In our dataset, we observe that there are average 590 heading objects and average 2300 physical objects in a document. Thus, the 2step-HELD model reduces around 75\% nodes for generating hierarchy, which causes average inquiry number reducing around 78\%. The 1step-HELD model consumes 355.18 seconds to process a document on average. In the 2step-HELD model, the heading recognition step consumes 1.2 seconds (2.8\% time) and the heading hierarchy generation step consumes 41.38 seconds (97.2\% time) on average, which means that the major time consumption comes from the heading hierarchy generation step. Thus, the 2step-HELD model obtains around 8.3x speedup ratio (from 355.18 to 42.58 seconds) in general compared with the 1step-HELD model.

In summary, to obtain higher effectiveness and efficiency, we extract headings explicitly in HELD model.

\begin{table*}
    \centering
    \caption{Exploring the performance of different features in put-or-skip model on the test set of the Chinese dataset.}
    \label{table:feature_chn}
    \scriptsize
    \begin{tabular}{lcccccccccccc}
        \hline
        \hline
        Model & $acc$ & $F1_1$ & $F1_2$ & $F1_3$ & $F1_4$ & $F1_5$ & $F1_6$ & $F1_7$ & $F1_8$ & $F1_9$ & $F1_{10}$ & $F1_{11}$ \\
        \hline
        2step-HELD (ta) & 0.9731 & 0.9892 & 0.9489 & 0.9784 & 0.9838 & 0.9778 & 0.9675 & 0.9449 & 0.9174 & 0.8521 & 0.8517 & 0.7059 \\
        - Parent & 0.9659 & 0.9889 & 0.9465 & 0.9729 & 0.9782 & 0.9686 & 0.9563 & 0.9410 & 0.9075 & 0.8481 & 0.9084 & 1.0000 \\
        - Siblings & 0.9402 & 0.9581 & 0.9131 & 0.9492 & 0.9564 & 0.9448 & 0.9334 & 0.9051 & 0.8471 & 0.6903 & 0.6575 & 0.3789 \\
        \hline
    \end{tabular}
\end{table*}

\begin{table*}
    \centering
    \caption{Exploring the performance of different features in put-or-skip model on the test set of the English dataset.}
    \label{table:feature_eng}
    \scriptsize
    \begin{tabular}{lcccccccccc}
        \hline
        Model & $acc$ & $F1_1$ & $F1_2$ & $F1_3$ & $F1_4$ & $F1_5$ & $F1_6$ & $F1_7$ & $F1_8$ & $F1_9$ \\
        \hline
        2step-HELD (ta) & 0.7301 & 0.8556 & 0.8380 & 0.8180 & 0.7211 & 0.6048 & 0.4807 & 0.3763 & 0.4258 & 0.4245 \\
        - Parent & 0.7007 & 0.7225 & 0.7608 & 0.7772 & 0.6902 & 0.5937 & 0.4908 & 0.4540 & 0.4643 & 0.4545 \\
        - Siblings & 0.6445 & 0.8242 & 0.7923 & 0.7614 & 0.6431 & 0.4642 & 0.3327 & 0.2609 & 0.1477 & 0.0000 \\
        \hline
    \end{tabular}
\end{table*}

\begin{table*}
    \centering
    \caption{Exploring the performance of beam search on the test set of the Chinese dataset.}
    \label{table:bs_chn}
    \scriptsize
    \begin{tabular}{lcccccccccccc}
        \hline
        Model & $acc$ & $F1_1$ & $F1_2$ & $F1_3$ & $F1_4$ & $F1_5$ & $F1_6$ & $F1_7$ & $F1_8$ & $F1_9$ & $F1_{10}$ & $F1_{11}$ \\
        \hline
        HELD (bs=1) & 0.9731 & 0.9892 & 0.9489 & 0.9784 & 0.9838 & 0.9778 & 0.9675 & 0.9449 & 0.9174 & 0.8521 & 0.8517 & 0.7059 \\
        HELD (bs=3) & 0.9736 & 0.9892 & 0.9489 & 0.9784 & 0.9838 & 0.9778 & 0.9685 & 0.9488 & 0.9179 & 0.8523 & 0.8734 & 1.0000 \\
        \hline
    \end{tabular}
\end{table*}

\begin{table*}
    \centering
    \caption{Exploring the performance of beam search on the test set of the English dataset.}
    \label{table:bs_eng}
    \scriptsize
    \begin{tabular}{lcccccccccc}
        \hline
        Model & $acc$ & $F1_1$ & $F1_2$ & $F1_3$ & $F1_4$ & $F1_5$ & $F1_6$ & $F1_7$ & $F1_8$ & $F1_9$ \\
        \hline
        HELD (bs=1) & 0.7301 & 0.8556 & 0.8380 & 0.8180 & 0.7211 & 0.6048 & 0.4807 & 0.3763 & 0.4258 & 0.4245 \\
        HELD (bs=3) & 0.7275 & 0.8277 & 0.8262 & 0.8148 & 0.7199 & 0.6053 & 0.4838 & 0.3775 & 0.4373 & 0.4813 \\
        \hline
    \end{tabular}
\end{table*}

\textbf{For RQ4}, we explore the tolerance to insertion errors of predecessor steps in HELD model, by evaluating $F1_k$ on the test set of the Chinese and English datasets. The results are shown in Table~\ref{table:baseline_chn}, Table~\ref{table:baseline_eng} and Table~\ref{table:baseline_arxiv}, respectively.

For comparison, we choose the best HELD model, which is 2step-HELD with the traversal-all method as the baseline. Note that the tolerance to insertion errors of predecessor steps is used in this model. Then, based on the best HELD model, we remove the tolerance to insertion errors of predecessor steps and obtain the last row in Table~\ref{table:baseline_chn} and Table~\ref{table:baseline_eng}. Clearly, after removing this process, the HELD model obtains node accuracy of 0.9725 and 0.7095 on the Chinese and English datasets, respectively. The results show that it obtains 0.0006 and 0.0206 decrease of node accuracy on the Chinese and English datasets, respectively. In other words, the tolerance to insertion errors obtains greater improvement on the English dataset, since there exist more insertion errors and adding the tolerance to insertion errors can make the HELD model insert nodes correctly based on some insertion errors of predecessor steps.

In summary, to obtain higher effectiveness, we add the tolerance to insertion errors of predecessor steps.

\textbf{For RQ5}, we design an ablation experiment to show the importance of contextual features in HELD model. Note that, according to Subsection~\ref{sec:3.3}, we consider that contextual features contain the previous siblings and the parent of the current node. For comparison, we choose the best HELD model, which is the 2step-HELD with the traversal-all method, as the baseline. Then, we remove the features of previous siblings and the parent in the put-or-skip module respectively to present the importance of another one. The experimental results are shown in Table~\ref{table:feature_chn} and Table~\ref{table:feature_eng}. Clearly, removing parent features obtains the node accuracy of 0.9659 and 0.7007 on the Chinese and English datasets, respectively, with the decrease in the node accuracy of 0.0072 and 0.0294, respectively. On the other hand, removing previous siblings' features obtains the node accuracy of 0.9659 and 0.7007 on the Chinese and English datasets, respectively, with the decrease in the node accuracy of 0.0330 and 0.3286, respectively. In other words, adding previous siblings' features can obtain prominent improvement on effectiveness for the put-or-skip module, since previous siblings often have the same format features and consecutive item numbers with the current node. Since the previous siblings and the parent features both obtain improvement on effectiveness, we use both of them in the put-or-skip module.

\textbf{For RQ6}, beam search is traditionally adopted for tree generation. In Subsection~\ref{sec:3.4}, we have introduced how to use beam search in the proposed HELD model. Here, to explore the effectiveness and efficiency of using beam search, we choose the best HELD model, which is the 2step-HELD with the traversal-all method and set beam size as 1 (using greedy search), as the baseline. Then, we set beam size as 3 for comparison and the experimental results on the Chinese and English datasets are shown in Table~\ref{table:bs_chn} and Table~\ref{table:bs_eng}, respectively. Clearly, setting beam size as 3 obtains the node accuracy of 0.9736 and 0.7275 on Chinese and English datasets, respectively. In other words, setting beam size as 3 will not prominently improve the accuracy and even decrease the accuracy on the English dataset. The reason is that the put-or-skip module can distinguish different possible positions apparently in inference, thus it does not need to expend several possible positions in each search step.

Meanwhile, we also count the processing time and \#inquiry for an average document compared with setting beam size with 1 and 3. The results show that generating logical hierarchy for each document consumes 42.58 seconds on average when setting beam search as 1 and consumes 413.78 seconds on average when setting beam search as 3. In other words, using beam search obtains about 10x decrease in efficiency.

In summary, since setting beam size as 1 achieves the best tradeoff between effectiveness and efficiency, we use the greedy search (without beam search) in the HELD model.

\begin{table*}[!htb]
    \centering
    \caption{Exploring the performance of document layout recognition.}
    \label{table:physical}
    \small
    \begin{tabular}{lccc}
        \hline
        Dataset & $p$ & $r$ & $F1$ \\
        \hline
        Chinese dataset & 0.9668 & 0.9666 & 0.9667 \\
        English dataset & 0.9734 & 0.9742 & 0.9738 \\
        \hline
    \end{tabular}
\end{table*}

\begin{table*}[!htb]
    \centering
    \caption{Comparing logical hierarchy based on the predicted and labeled document layout in the Chinese dataset.}
    \label{table:from_element_zh}
    \scriptsize
    \begin{tabular}{lcccccccccccc}
        \hline
        Level & Total & 1 & 2 & 3 & 4 & 5 & 6 & 7 & 8 & 9 & 10 & 11 \\
        \hline
        Labeled Layout & 0.9731 & 0.9892 & 0.9489 & 0.9784 & 0.9838 & 0.9778 & 0.9675 & 0.9449 & 0.9174 & 0.8521 & 0.8517 & 0.7059 \\
        Predicted Layout & 0.9576 & 0.9646 & 0.9273 & 0.9737 & 0.9623 & 0.9608 & 0.9535 & 0.9423 & 0.9103 & 0.8466 & 0.8345 & 0.6667 \\
        \hline
    \end{tabular}
\end{table*}

\begin{table*}[!htb]
    \centering
    \caption{Comparing logical hierarchy based on the predicted and labeled document layout in the English dataset.}
    \label{table:from_element_en}
    \scriptsize
    \begin{tabular}{lcccccccccc}
        \hline
        Level & Total & 1 & 2 & 3 & 4 & 5 & 6 & 7 & 8 & 9 \\
        \hline
        Labeled Layout & 0.7301 & 0.8556 & 0.8380 & 0.8180 & 0.7211 & 0.6048 & 0.4807 & 0.3763 & 0.4258 & 0.4245 \\
        Predicted Layout & 0.7238 & 0.8491 & 0.8315 & 0.8051 & 0.7110 & 0.5939 & 0.4586 & 0.3605 & 0.3976 & 0.4138 \\
        \hline
    \end{tabular}
\end{table*}

\textbf{For RQ7}, we aim to explore the influence of noise in document layout recognition.
As mentioned previously, we adopt a commercial product, PDFLux\footnote{\url{https://pdflux.com/}. Last visited in 2021/04/28.}, for document layout recognition, which detects physical objects (e.g. paragraphs, tables, graphs) on each document page.
Since the document physical objects are labeled by annotations in each dataset, we can evaluate the performance of PDFLux.
Here, we use a rigorous metric. First, we define the exact match of a predicted object if it detects the exact region without missing any text or containing any redundant text outside objects compared with the ground-truth object. Then, we can calculate the precision, recall and F1 value of the exact-matched physical objects.

The results of Chinese and English datasets are shown in Table~\ref{table:physical}.
PDFLux obtains 0.9667 and 0.9738 F1 in the Chinese and English datasets, respectively.
In other words, around 97\% predicted physical objects exactly match the ground-truth objects.

Next, based on the predicted physical objects, we can use the HELD model to recognize the logical hierarchy of each document.
Thus, we evaluate the logical hierarchy based on the predicted physical objects as shown in Table~\ref{table:from_element_zh} and Table~\ref{table:from_element_en}.
Compared with recognizing logical hierarchy based on the labeled physical objects, recognizing logical hierarchy based on predicted physical objects obtains 0.0155 (from 0.9731 to 0.9576) and 0.0063 (from 0.7301 to 0.7238) decrease of accuracy in the Chinese and English dataset.
That is to say, the noise of document layout recognition has limited influence on the discovery of logical document hierarchy.

\subsection{Experimental Summary}

In summary, HELD model greatly outperforms three baseline models on effectiveness. Meanwhile, since extracting headings explicitly obtains great improvement on effectiveness and efficiency, we choose the 2step-HELD model. To obtain higher generalization ability, we add the tolerance to insertion errors in the predecessor step. To achieve the tradeoff between effectiveness and efficiency, we choose root-to-leaf traversal order. Additionally, leaf-to-root traversal method can be used if the requirement of efficiency is much higher in the real-world production.

\begin{figure*}[!htb]
    \centering
    \subfigure[]{\label{fig:case_study}
    \includegraphics[width=5.5in]{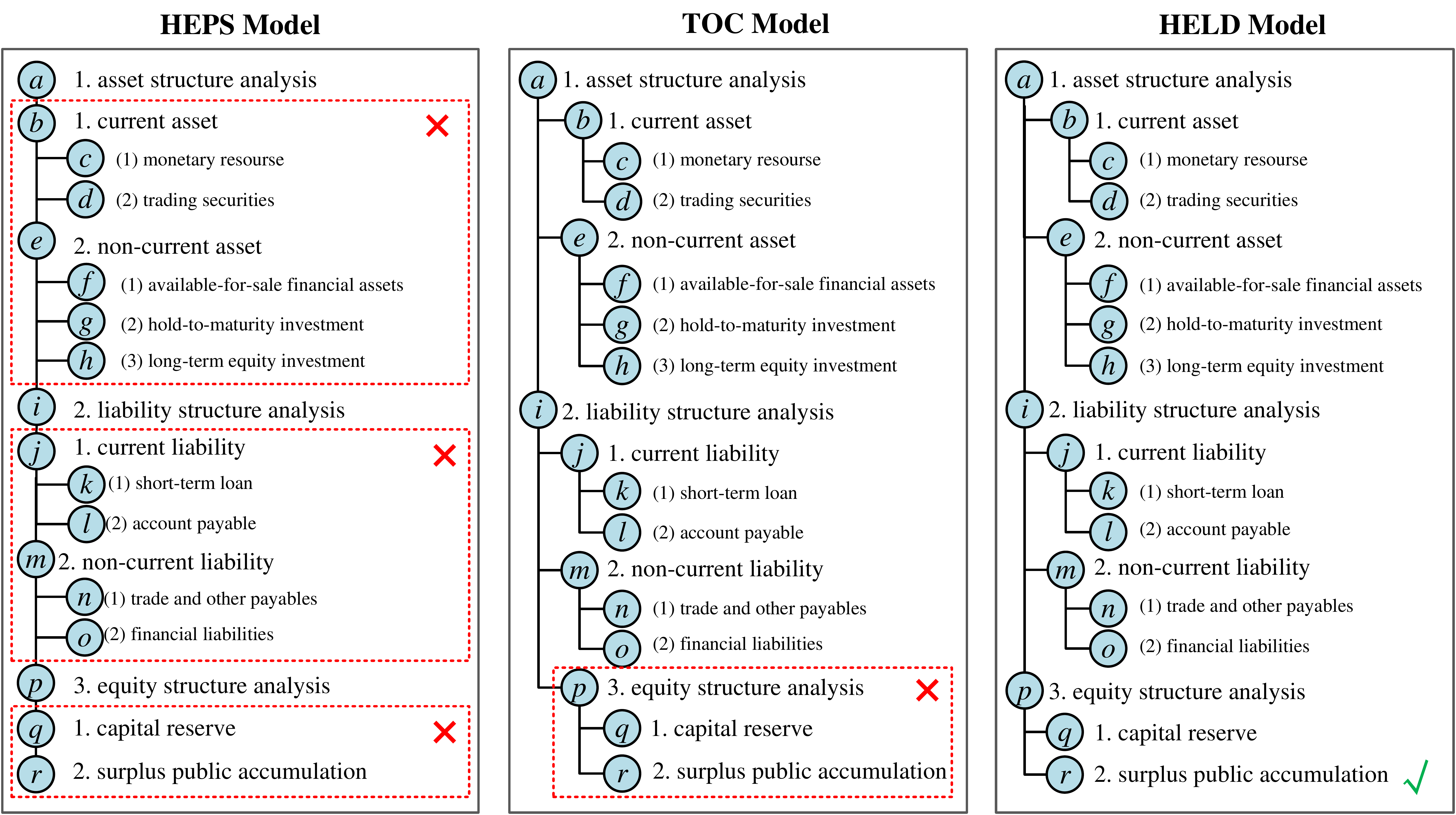}}
    \subfigure[]{\label{fig:limitation}
    \includegraphics[width=4.8in]{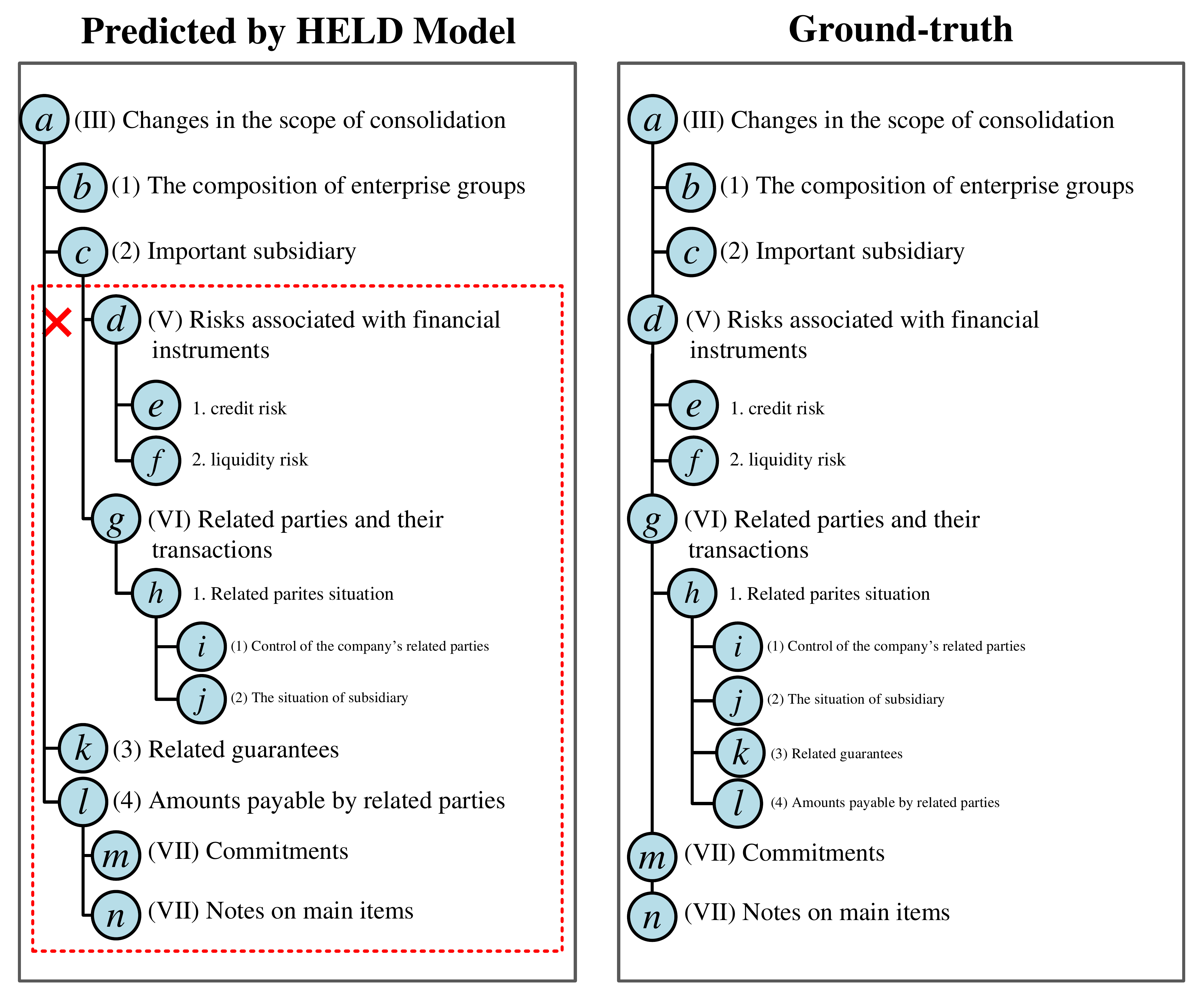}}
    \caption{Case Studies. (a) Comparing HEPS, TOC and HELD model. (b) An example to show the limitation of the HELD model.}
    \label{fig:case_studies}
\end{figure*}

\subsection{Case Studies}

In this subsection, we first show an example to compare the HELD model, the HEPS model (rule-based) and the TOC model (sequence labeling-based) models.
As shown in Fig.~\ref{fig:case_study}, the difficulty of this example is that high-level and low-level headings use the same pattern (the pattern starts with a number and a punctuation, like ``1. $\cdots$'', ``2.$\cdots$''), so that it is hard to decide the true level of each heading.
The HEPS model predicts the incorrect position of ``2. non-current asset'' with its descendants and the TOC model also predicts incorrectly for the last three headings. However, the HELD model correctly predicts hierarchy via text and contextual information, since it considers the parent and siblings information simultaneously for each heading.

Then, we use another example to show the limitation of the proposed HELD model.
For example in Fig.~\ref{fig:limitation}, the ground-truth hierarchy is shown on the right part.
Note that, the next sibling of heading $a$, ``($\textnormal{\uppercase\expandafter{\romannumeral3}}$) Changes $\cdots$'', is heading $d$, ``($\textnormal{\uppercase\expandafter{\romannumeral3}}$) Risks $\cdots$''. The heading $($\textnormal{\uppercase\expandafter{\romannumeral4}}$) \cdots$ is omitted for some reasons, like the errors of heading detection step or the errors of original document editing.
In this scenario, the put-or-skip module predicts heading $d$ as the child of heading $c$ by mistake, which causes that all the subsequent headings are predicted incorrectly.
Thus, the HELD model has a limitation. It is hard to correctly predict the headings in the lower level once a heading in the higher level is predicted incorrectly.
The main reason may be that the hierarchy generation process is a greedy search process in the HELD model.
In the future, we aim to tackle this problem via the Monte Carlo Tree Search technique.


\section{Downstream Application}

We further explore how the logical document hierarchy can be leveraged in a downstream application of passage retrieval.  Professionals usually need to retrieve the relevant passages in a document with hundreds of pages. Here, we define ``passage'' as paragraph, table, figure and so on in the document. Clearly, this task can be formulated as a learning-to-rank problem for all the content passages within a document. We show that the features extracted from the document hierarchy can significantly improve retrieval performance.

Generally, for any passage corresponding to a node in the logical tree, the features on the path from this node to the root may help on this task.  Hence, besides the traditional BM25 feature four more features are extracted based on the document hierarchy. Specifically, ``BM25AncMax'' is the maximum BM25 score among the ancestor nodes of a given passage. The heading of a section usually contains the general description of its subsections. If the section heading hits the query keyword, the passages in this section are more likely to be relevant. ``SameWordAnc'' is the number of the same words between a given passage and its ancestors. We merge all the ancestors of the passage into one text and calculate the intersection number of words between the passage and the merged text. The ancestor nodes contain a general summary about the content under them, so the passage that contains more intersecting words has higher importance. ``Pos'' and ``PosRatio'' are the absolute and relative indexes of a given passage among its siblings. Note that, $PosRatio=\frac{Pos}{c}$, where $c$ is the number of siblings of the passage. These two features point out the positional information of children, where the first and last child often provide an integrated description that may have a higher rank. Based on these features, Gradient Boosting Decision Tree (GBDT)~\cite{friedman2001greedy} is adopted to rank the passages.

\vspace{2mm}
\begin{center}
    \scriptsize
    \captionof{table}{The results of passage retrieval.}\label{table:app}
	\begin{tabular}{lcccc}
        \hline
        \multicolumn{1}{c}{\multirow{2}{*}{Adding Features}} & \multirow{2}{*}{$mAP$} & \multicolumn{3}{c}{$recall@k$} \\
        \multicolumn{1}{c}{} &  & $k$=1 & $k$=5 & $k$=10 \\
        \hline
		Only BM25 & 0.149 & 0.083 & 0.296 & 0.412 \\
        BM25 + BM25AncMax & 0.269 & 0.184 & 0.376 & 0.465 \\
        BM25 + SameWordAnc & 0.223 & 0.126 & 0.336 & 0.487 \\
        BM25 + Pos & 0.254 & 0.165 & 0.403 & 0.519 \\
        BM25 + PosRatio & 0.219 & 0.127 & 0.335 & 0.478 \\
        BM25 + All Four Features & 0.338 & 0.218 & 0.471 & 0.576 \\
        \hline
	\end{tabular}
\end{center}
\vspace{4mm}

The dataset contains 110 IPO prospectus in the Chinese market and a set of 138 queries are applied to each document. We spilt the dataset by queries with 88 queries for training and 50 queries for testing. Table~\ref{table:app} shows the measures of mAP and recall on this testing for different sets of used features. The baseline model is to use only the BM25 feature. Then, each of the four new features is combined to get another four baseline models. Finally, we combine the four features with BM25 to get the final model. The experimental results show that each of the four features can improve the retrieval performance and the BM25AncMax and Pos feature get relative prominent improvement. With all the four features together, we obtain $0.189$ increase on mAP and $0.135$ increase on recall@1. In conclusion, logical document hierarchy can be employed to significantly improve the performance of the downstream passage retrieval task.

It is worth mentioning that the proposed four new features heavily depend on the path from a node to the root. Any errors in this path might seriously worsen this application. Therefore, we argue again that the proposed measure which checks the path from any node to the root is more reasonable.

\section{Conclusion}

In this paper, we conducted a systematic study on the task of extracting logical document hierarchy from long documents in terms of methods, evaluations, and applications. We showed that the proposed HELD model with the root-to-leaf traversal order and explicit heading extraction is suitable to achieve the tradeoff between effectiveness and efficiency. Furthermore, we demonstrated that the downstream passage retrieval task significantly benefits from the extracted tree. We also argued that this proposed new measure should be adopted in future studies of this task.

\vspace{2mm}


{
\footnotesize
\itemsep=-3pt plus.2pt minus.2pt
\baselineskip=14pt plus.2pt minus.2pt
\bibliographystyle{JCSTbib}
\bibliography{reference}
}

\end{multicols}

\end{document}